\documentclass[preprint,10pt]{elsarticle}
\usepackage[english]{babel}
\usepackage{latexsym,amsmath,amssymb,amsbsy,amstext,amscd,amsfonts,amsthm}
\usepackage{graphics,graphicx}
\usepackage{color}
\usepackage[dvipsnames]{xcolor}
\usepackage{tabularx}
\usepackage{multirow}
\usepackage{booktabs}
\usepackage{float}
\theoremstyle{definition}
\newtheorem{remark}{Remark}
\usepackage{array}
\usepackage{amssymb}
\newcommand{\beq}{\begin{equation}}
\newcommand{\eeq}{\end{equation}}

\newcommand{\beqar}{\begin{eqnarray}}
\newcommand{\eeqar}{\end{eqnarray}}
\newcommand{\bit}{\begin{itemize}}
\newcommand{\eit}{\end{itemize}}
\newcommand{\benum}{\begin{enumerate}}
\newcommand{\eenum}{\end{enumerate}}
\newcommand{\barr}{\begin{array}}
\newcommand{\earr}{\end{array}}

\def\XXint#1#2#3{{\setbox0=\hbox{$#1{#2#3}{\int}$}
   \vcenter{\hbox{$#2#3$}}\kern-.5\wd0}}

\def\b0{\mbox{\boldmath $0$}}

\def\f0{\ensuremath{\mathbb{O}}}

\begin{document}

\begin{frontmatter}

%\maketitle
\title{An efficient algorithm of solution for the flow of generalized Newtonian fluid in channels of simple geometries}

\author{Michal Wrobel }
\ead{wrobel.michal@ucy.ac.cy}
\address{\it  Department of Civil and Environmental Engineering, University of Cyprus,
\\ {\it 75 Kallipoleos Street, 1678 Nicosia, Cyprus}}

\begin{abstract}
In this paper a problem of  stationary flow of generalized Newtonian fluid in a thin channel is considered. An efficient algorithm of solution is proposed that includes a  flexible procedure for a continuous approximation of the apparent viscosity by means of elementary functions combined with analytical integration of the governing equations. The algorithm can be easily adapted to circular or elliptic conduits. The accuracy and efficiency of computations are analyzed using an example of the Carreau fluid. The proposed computational scheme proves to be highly efficient and versatile  providing excellent accuracy of solution at a very low computational cost. 
\end{abstract}


\end{frontmatter}

\section{Introduction}

The problem of a laminar flow of non-Newtonian fluid in straight conduits and slots is commonly encountered in nature and technology. The need for mathematical modelling of such a problem arises for example in geology \cite{Rubin_1995,de_Castro_2017}, reservoir engineering \cite{Perkowska_2016,Peck_2018_1,Peck_2018_2,Osiptsov_2017}, bioengineering \cite{Rubenstein_2015,Brujan_2011} or chemical engineering \cite{Abulencia_2009}. In many of these applications it is vital to obtain a relation between the pressure gradient and the velocity profile or the volumetric flow rate for given rheological properties of the fluid. Unfortunately, the closed form solutions   exist only for few rheological models \cite{Bird_1987,Carreau_1997,Housiadas_2015}.  To circumvent this problem the empirical relations are used instead \cite{Kozicki_1966} or numerical mesh-based techniques  are employed \cite{KimN_2019,Kwon_2018}. Nevertheless, in the former case the question of accuracy of such estimations often arises. When using numerical computations,  the efficiency and stability issues also become crucial, especially in those problems where multiple iterative estimations of the  velocity and the flow rate are needed for large numbers of points (e.g. in the hydraulic fracture problem).

The constitutive relations describing rheological properties of fluids can be of different forms \cite{Perlacova_2015,Kim_2018}. One of the most popular among them  is  a concept of the generalized Newtonian fluid \cite{Bird_1987} in which the shear stress is an explicit function of the shear rate. Within this formulation fluids such as Carreau and Cross \cite{Bird_1987} are classified. These two models are widely used in describing  rheological proprieties of biological fluids and polymeric liquids. Thus, they are frequently employed in simulations of flow in blood arteries \cite{Akbar_ 2014,Gholipour_2018} or rheometric measurements \cite{Busch_2018}. Moreover, their ability to capture asymptotic values of apparent viscosity at high and low shear rates combined with shear-thinning behaviour in the interim makes them very useful in describing fracturing fluids in reservoir stimulation \cite{Lecampion_2018}, especially in the cases where the pure power-law model cannot be used \cite{Wrobel_2017,Wrobel_2018}. Unfortunately, no closed-form analytical solutions for  velocity and fluid flow rate are available for the Carreau and Cross models, which constitutes a serious hindrance in their implementation. 

One of the possible ways to overcome this difficulty is to use a substitute expression for apparent viscosity that preserves  basic features of the original law and simultaneously is simple enough to allow analytical integration. These requirements  are satisfied by the truncated power-law model \cite{Bird_1987}. By accounting for high and low shear rate cut-off viscosities this model constitutes simple regularization of the pure power-law. The truncated power-law model was used in \cite{Lavrov_2015}  to simulate  fluid flow through a flat channel (slit flow). The author derived analytical formulae for both, the fluid velocity and the fluid flow rate. The results were compared with a numerical solution computed for the Carreau fluid and  analytical relations available for the pure power-law model. As anticipated, the truncated power law yields much better resemblance of results obtained for the Carreu model than the pure power-law variant over a wide range of  pressure gradients. However, the quality of such an approximation for some values of the shear rates  may not be sufficient for practical applications. A slight improvement to the truncated power-law model can be found in \cite{Ruschak_2013} where the authors introduced an additional local power-law approximation around the  points of intersection between the plateau visocisites and the intermediate power-law characteristics (the Carreau-Yasuda model is considered as a reference example).

Another attempt to facilitate implementation of the generalized Newtonian fluids, including the Carreau and Cross models, was made in \cite{Sochi_2014}. The proposed method is based on the Euler-Lagrange variational principle. In the cases where analytical integration of the equations derived from the variational principle is impossible, the method is  combined with numerical integration schemes to obtain the fluid velocity and the  fluid flow rate. In particular, for the Carreau and Cross fluids, one has to solve non-linear algebraic equations for the values of  shear rates across the channel radius. Next, the obtained shear rate profile is  integrated numerically for velocity and  flux.  As this is a mesh based method, sufficient mesh density is required for convergence. Moreover, presence of the hypergeometric Gauss functions in the equations for the shear rates can reduce efficiency and accuracy of computations as these functions may not converge satisfactorily when computed numerically. Finally, as noted by the author, the method fails for high yield stress fluids.

The semi-analytical solutions for the fluid flow rate in a circular pipe and a flat channel were derived in \cite{Sochi_2015} for the Carreau and Cross fluids. These solutions were obtained in the framework of the Weissenberg-Rabinowitsch-Mooney-Schofield method. This approach was extended in \cite{Kim_2018} to compute the fluid velocity in a circular pipe (in \cite{Kim_2019} the author employs the same methodology for the case of viscoplastic fluids). Generally, computation of the fluid velocity and flow rates  includes numerical solving of non-linear algebraic equations to find fluid shear rates and their subsequent application in respective analytical formulae. Depending on the channel geometry and the rheological model, the pertinent analytical relations can include the hypergeometric Gauss functions, which results in computational problems mentioned  above. Furthermore, in those variants of solution where only elementary functions are used to compute the fluid velocity and flux, there appear singularities of the respective expressions for the fluid behaviour index equal to 1/3. Nevertheless, the authors did not comment on this fact. The semi-analytical solutions of the described type are available only for some rheological models where  indefinite integrals of certain constitutive relations are available. In other cases of generalized Newtonian fluids  numerical integration should be employed. 

In this paper we introduce an efficient computational scheme that can be used to obtain the values of fluid velocity and flow rate in the conduits of simple geometries (slit flow, circular or elliptic pipe) for an arbitrary generalized Newtonian fluid.  We propose a flexible procedure for a continuous approximation of the apparent viscosity with any desired level of accuracy. The approximation model is based on elementary functions. Thus, the respective governing equations can be integrated analytically to produce a solution expressed by elementary functions as well. The new algorithm is very effective since it does not require usage of special functions, iterative computations or solving systems of algebraic equations. 

The paper is structured in the following way. In Section \ref{gov_rel} we define the problem geometry and the governing equations. Section \ref{const_rel} is devoted to description of constitutive relations that will be used to demonstrate  performance of the new algorithm.  In section \ref{TPL_sol} the truncated power-law problem is analyzed, motivated by the fact that it  is very instructive in understanding the essentials of the solution proposed for the generalized Newtonian fluid. In Section \ref{apr_sec} the computational algorithm for the generalized Newtonian fluid is formulated. The  accuracy and efficiency of computations by the proposed scheme are investigated on the example of Carreau fluid in Section \ref{num_res}. Finally, some key findings of the paper are summarized in Section \ref{disc}.

\section{The general relations}
\label{gov_rel}
\begin{figure}[htb!]
\begin{center}
\includegraphics[scale=0.8]{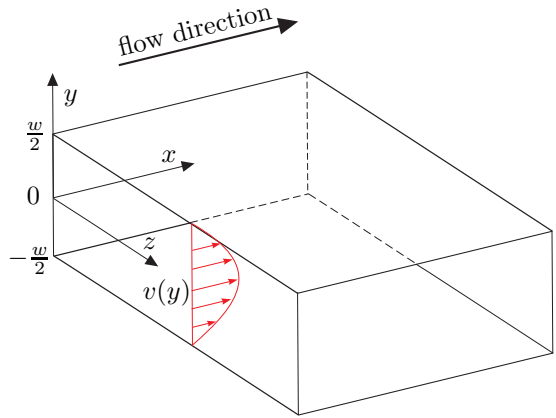}
\put(-185,120){$y$}
\put(-148,97){$x$}
\put(-155,64){$z$}
\put(-155,45){$v(y)$}
\put(-200,105){$\frac{w}{2}$}
\put(-206,58){$-\frac{w}{2}$}
\put(-199,81){$0$}
\put(-165,137){\rotatebox{14}{\text{flow direction}}}
\caption{Fluid flow in a flat channel. }
\label{flat_c}
\end{center}
\end{figure}

Let us consider a problem of a stationary fully developed laminar flow of incompressible fluid in a flat channel (slit flow) schematically shown in Fig. \ref{flat_c}. The following governing equation can be derived on the assumption of a unidirectional fluid movement in the absence of  mass forces (see e.g. \cite{Kim_2019,Perkowska_Phd}):
\begin{equation}
\label{gov_ode}
\frac{\text{d}}{\text{d} y}\tau (\dot \gamma)=\frac{\text{d} p}{\text{d}x}, \quad y \in\left[-\frac{w}{2},\frac{w}{2}\right]
\end{equation}
where $\tau$ is a shear stress in the direction of flow which in the general case depends on the shear rate $\dot \gamma$:
\begin{equation}
\label{gam_def}
\dot \gamma=\frac{\text{d} v}{\text{d}y},
\end{equation}
with $v=v(y)$ being the fluid velocity. The pressure gradient $\frac{\text{d} p}{\text{d}x}<0$ is assumed constant along the channel length.  We adopt here a definition of the generalized Newtonian fluid \cite{Bird_1987} which describes the shear stress as:
\begin{equation}
\label{GNF}
\tau(\dot \gamma)=\eta_\text{a}^{(\text{g})}\dot \gamma, 
\end{equation}
where $\eta_\text{a}^{(\text{g})}=\eta_\text{a}^{(\text{g})}(\dot \gamma)$ is the so called apparent viscosity\footnote{The superscript `$\text{g}$' will be used throughout this paper to denote the general form of apparent viscosity.}.

 As the problem is symmetric with respect to the plane $y=0$, we consider only the upper part of the channel with the following boundary conditions:
\begin{equation}
\label{BCs}
\dot \gamma (0)=0, \quad v \left(\frac{w}{2}\right)=0.
\end{equation}
Note also that:
\begin{equation}
\label{gamma_neg}
\dot \gamma \leq 0, \quad y \in \left[0,\frac{w}{2}\right].
\end{equation}
The average fluid flow rate through the channel (per unit length in the $z$-direction) can be computed as:
\begin{equation}
\label{Q_def}
Q=\int_{-w/2}^{w/2}v(y)dy=2\int_{0}^{w/2}v(y)dy.
\end{equation}

In the following we propose a simple and efficient method for numerical computation of the fluid flow rate and velocity for fluids whose rheological properties are described by the general relation \eqref{GNF}.

\section{The constitutive relations}
\label{const_rel}

Let us describe here three examples of constitutive relations that have the form of the generalized Newtonian fluid: i) the power-law fluid, ii) the Carreau fluid, iii) the truncated power-law fluid. These examples will be used later on to explain the construction of the general algorithm and verify its performance. Moreover, the analytical solutions for the pure power law and truncated power law will be employed in \ref{ap_num_sol} to estimate the accuracy of the constructed numerical reference solution for the Carreau problem.

The simplest model capable of describing the non-Newtonian behaviour of a fluid is the power-law model \cite{Bird_1987,Gholipour_2018_1}:
\begin{equation}
\label{PL_eta}
\eta_\text{a}= C |\dot \gamma|^{n-1},
\end{equation}
where $C$ is the consistency index, and $n$ stands for the fluid behaviour index. For $n<1$ one obtains the shear thinning properties, while $n>1$ gives the shear thickening characteristic. Unfortunately, the power-law model produces unrealistic results for both, the high and the low shear rate values \cite{Lecampion_2018,Lavrov_2015}. In the case of the shear-thinning fluids the power-law model underestimates the apparent viscosity for large  $|\dot \gamma|$ and overestimates it for small shear rates. For the shear-thickening fluids a reverse relation holds. 

More accurate reproduction of the real fluid rheology is provided by four parameter rheological models such as Carreau or Cross. In this paper we will use an example of the  former which is described by the formula:

\begin{equation}
\label{carreau_def}
\eta_\text{a}=\eta_\infty+(\eta_0-\eta_\infty)\left[1+(\lambda\dot\gamma)^2 \right]^{\frac{n-1}{2}},
\end{equation}
where $\eta_0$ and $\eta_\infty$ are the limiting viscosities for low and high shear rates, respectively, while $n$ and $\lambda$ are the fitting parameters. Nevertheless, the Carreau model is problematic in implementation as it does not allow for analytical integration of the governing equation \eqref{gov_ode} to obtain the velocity profile and the fluid flow rate.

\begin{figure}[htb!]
\begin{center}
\includegraphics[scale=0.5]{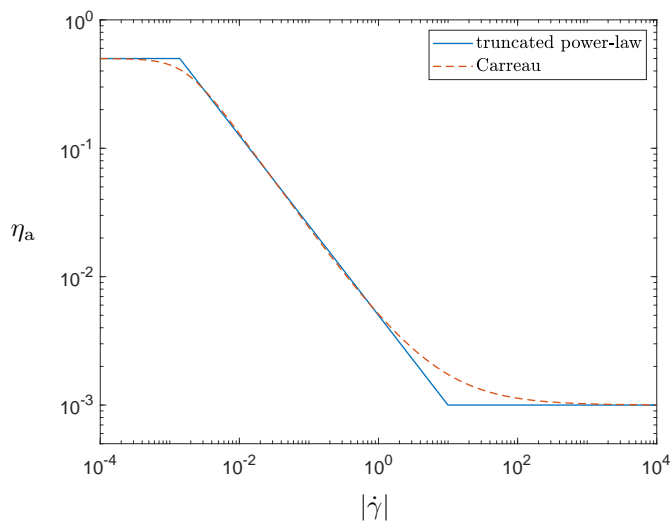}
\put(-138,-5){$|\dot \gamma|$}
\put(-270,100){$\eta_\text{a}$}
\caption{The apparent viscosities for the Carreau and the truncated power-law models for the data from Table \ref{par_table}.}
\label{Tpl_C_graph}
\end{center}
\end{figure}

To remedy this drawback and simultaneously to enable analytical solution of the problem which accounts for the limiting cut-off viscosities, the truncated power-law model was introduced \cite{Bird_1987}. It is expressed by the relation:
\begin{equation}
\label{TP_def}
\eta_\text{a}=
  \begin{cases}
		\eta_0       & \quad \text{for } \quad |\dot \gamma |<|\dot \gamma_1|,\\
    C |\dot \gamma|^{n-1}       & \quad \text{for} \quad |\dot \gamma_1|<|\dot \gamma|<|\dot \gamma_2|,\\
		
    \eta_\infty  & \quad \text{for } \quad |\dot \gamma|>|\dot \gamma_2|.
  \end{cases}
\end{equation}
 In order to best  fit the Carreau characteristic, the value of $n$ can be taken different from the one used in \eqref{carreau_def} \cite{Lecampion_2018,Lavrov_2015}. The limiting values of shear rates for which the cut-off viscosities are employed are:
\begin{equation}
\label{gamma_lim}
|\dot \gamma_1|=\left(\frac{C}{\eta_0}\right)^{1/(1-n)}, \quad |\dot \gamma_2|=\left(\frac{C}{\eta_\infty}\right)^{1/(1-n)}.
\end{equation}
In our analysis we will use the parameters of the Carreau and the power-law models provided in \cite{Lavrov_2015}. They are summarized in Table \ref{par_table}. The graphical comparison of respective viscosity curves is shown in Fig. \ref{Tpl_C_graph}.

\begin{table}
\begin{center}
\begin{tabular}{ |p{4cm}||p{4cm}|p{4cm}|  }
 \hline
Rheological model& Parameter &Value\\
 \hline
 \hline
Carreau   & $\eta_0$ [Pa $\cdot$ s]    &0.5\\
 &   $\eta_\infty$ [Pa $\cdot$ s]    & 0.001\\
 &$n$ & 0.25\\
 &$\lambda$ [s] & 600\\
 \hline
 Truncated power-law&   $\eta_0$ [Pa $\cdot$ s] & 0.5\\
 & $\eta_\infty$ [Pa $\cdot$ s]   & 0.001\\
 & $n$   & 0.3\\
 & $C$ [Pa $\cdot$ s$^{n}$] &0.005 \\
 \hline
\end{tabular}
\end{center}
\caption{Parameters of the Carreau and the truncated power-law models according to \cite{Lavrov_2015}.}
\label{par_table}
\end{table}

\section{Solution to the truncated power-law model}
\label{TPL_sol}

\begin{figure}[htb!]
\begin{center}
\includegraphics[scale=0.6]{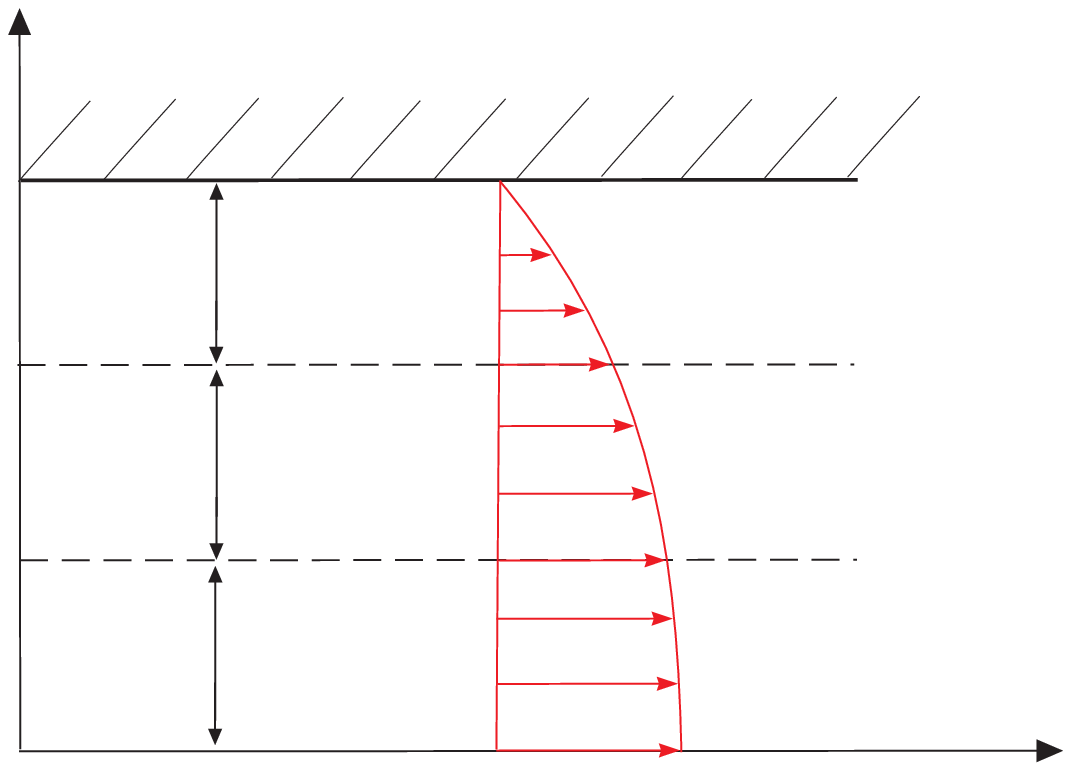}
\put(-13,7){$x$}
\put(-175,120){$y$}
\put(-190,0){$0$}
\put(-193,33){$\delta_1$}
\put(-213,67){$\delta_1+\delta_2$}
\put(-193,100){$\frac{w}{2}$}
\put(-160,15){\rotatebox{90}{$\delta_1$}}
\put(-160,47){\rotatebox{90}{$\delta_2$}}
\put(-160,80){\rotatebox{90}{$\delta_3$}}
\put(-65,47){$v(y)$}
\put(-135,40){$v_1,  \dot \gamma_1$}
\put(-135,74){$v_2,  \dot \gamma_2$}
\caption{The channel cross section and velocity profile. Only the upper symmetrical part is shown.}
\label{channel}
\end{center}
\end{figure}

A solution to the problem of the truncated power-law was given in \cite{Lavrov_2015}. The author provided the formulae for the fluid velocity and the average fluid flow rate obtained by the analytical integration of the governing ODE \eqref{gov_ode} with the apparent viscosity defined by \eqref{TP_def}. It is important to note that, depending on the values of the channel height, $w$, and pressure gradient, $dp/dx$, up to three distinct layers can appear across each of the slit symmetrical parts (see Fig. \ref{channel}):
\begin{itemize}
\item{The low shear rate domain,  located in the very core of the flow. The Newtonian-type behaviour of the fluid holds here. This layer is always present regardless of the values of $w$ and $dp/dx$. Its thickness is defined by $\delta_1$:
\begin{equation}
\label{delta_1_def}
\delta_1=-\left(\frac{dp}{dx} \right)^{-1}\eta_0^{\frac{n}{n-1}}C^{\frac{1}{1-n}}.
\end{equation}
Naturally, if $\delta_1\geq w/2$ then the layer thickness is limited by the channel height. }
\item{The intermediate shear rate domain where the power-law behaviour of the fluid is observed. Its thickness described by $\delta_2$ is:
\begin{equation}
\label{delta_2_def}
\delta_2=-\left(\frac{dp}{dx} \right)^{-1}C^{\frac{1}{1-n}}\left(\eta_\infty^{\frac{n}{n-1}}- \eta_0^{\frac{n}{n-1}}\right).
\end{equation} 
Again, if $\delta_1+\delta_2\geq w/2$ then the upper boundary of the layer is defined by the channel wall. If across the whole slit height $|\dot \gamma|<|\dot \gamma_1|$ then the layer does not appear at all.}
\item{The high shear rate domain that covers this part of the channel for which  $|\dot \gamma|>|\dot \gamma_2|$. Its thickness is described by $\delta_3$:
\begin{equation}
\label{delta_3_def}
\delta_3=\frac{w}{2}-\delta_1-\delta_2.
\end{equation}
This layer, if appears, is adjacent to the channel walls.  The Newtonian model of the fluid is valid here.}
\end{itemize}

Respective layers are depicted in Fig. \ref{channel}. Note that for a constant channel height, if the pressure gradient magnitude is increased, the first two layers are reduced with the third one growing. 

In \cite{Lavrov_2015} the author provides the formulae for the fluid velocity in respective layers. Below, we will reproduce the pertinent expressions in a different convention. First, let us accept the following notation for the case where all three shear rate layers are present:
\begin{equation}
\label{v_layers}
v(y)=
  \begin{cases}
		v^{(1)}(y)      & \quad \text{for } \quad 0\leq y<\delta_1,\\
    v^{(2)}(y)      & \quad \text{for} \quad \delta_1\leq y<\delta_1+\delta_2,\\
		
    v^{(3)}(y)  & \quad \text{for } \quad \delta_1+\delta_2 \leq y \leq w/2.
  \end{cases}
\end{equation}
The corresponding velocities are obtained by analytical integration of the governing equations with respective boundary conditions:
\begin{equation}
\label{v_I_def}
v^{(1)}(y)=v_1-\frac{1}{2\eta_0}\frac{dp}{dx}\left(\delta_1^2-y^2 \right),
\end{equation}
\begin{equation}
\label{v_II_def}
v^{(2)}(y)=v_2+\frac{n}{n+1}\left( -\frac{1}{C}\frac{dp}{dx}\right)^\frac{1}{n}\left[ (\delta_1+\delta_2)^\frac{n+1}{n}-y^\frac{n+1}{n} \right],
\end{equation}
\begin{equation}
\label{v_III_def}
v^{(3)}(y)=-\frac{1}{2\eta_\infty}\frac{dp}{dx}\left(\frac{w^2}{4}-y^2 \right).
\end{equation}
The interface values of the velocity can be easily computed as: $v_2=v^{(3)}(\delta_1+\delta_2)$ and $v_1=v^{(2)}(\delta_1)$, which provides continuity of $v$.

When for the predefined values of $w$ and $dp/dx$ only two shear rate layers exist, the solution can be recreated from equations \eqref{v_I_def} -- \eqref{v_II_def} by setting $v_2=0$ and $\delta_1+\delta_2=w/2$. Similarly, if there is only the low shear rate layer, the velocity is given by equation \eqref{v_I_def} for $v_1=0$ and $\delta_1=w/2$.

The average fluid flow rate can be  computed in an analytical way by employing the respective velocity representation in \eqref{Q_def}. The corresponding formulae are:
\begin{itemize}
\item{for the case of a single (low shear rate) layer
\begin{equation}
\label{Q_1}
Q=-\frac{1}{12\eta_0}\frac{dp}{dx}w^3,
\end{equation}}
\item{for the case of two (low and intermediate shear rate) layers
\begin{equation}
\label{Q_2}
Q=\frac{2(1-n)}{3(1+2n)}\left( \frac{dp}{dx}\right)^{-2}C^\frac{3}{1-n}\eta_0^\frac{2n+1}{n-1}+\frac{2n}{2n+1}\left( -\frac{1}{C}\frac{dp}{dx}\right)^\frac{1}{n}\left(\frac{w}{2} \right)^\frac{2n+1}{n},
\end{equation}}
\item{for the case of three (low, intermediate and high shear rate) layers
\begin{equation}
\label{Q_3}
Q=-\frac{1}{12\eta_\infty}\frac{dp}{dx}w^3+\frac{2(1-n)}{3(1+2n)}\left(\frac{dp}{dx}\right)^{-2}C^\frac{3}{1-n}\left(\eta_0^\frac{2n+1}{n-1}-\eta_\infty^\frac{2n+1}{n-1} \right).
\end{equation}}
\end{itemize}

As shown in \cite{Lavrov_2015}, the velocity profile computed with the truncated power law  imitates the results obtained for the Carreau law appreciably better than the one using the classic power-law rheology. However, in certain ranges of shear rates the accuracy of this approximation is not satisfactory and may  not be sufficient for practical purposes \cite{Ruschak_2013}. 

\section{Approximate solution for the generalized Newtonian fluid }
\label{apr_sec}

Below we present a direct extension of the idea of the truncated power-law model that can be used to approximate  other rheological characteristics (including those given by experimental data) with an arbitrary accuracy. Let us assume that the generalized apparent viscosity, $\eta_\text{a}^{(\text{g})}$, can be approximated in a continuous manner by the following expression:
\begin{equation}
\label{eta_ap}
\eta_\text{a}=
  \begin{cases}
		\eta_0       & \quad \text{for } \quad |\dot \gamma |<|\dot \gamma_1|,\\
    C_j |\dot \gamma|^{n_j-1}       & \quad \text{for} \quad |\dot \gamma_j|<|\dot \gamma|<|\dot \gamma_{j+1}| \quad j=1,...,N-1,\\
		
    \eta_\infty  & \quad \text{for } \quad |\dot \gamma|>|\dot \gamma_N|,
  \end{cases}
\end{equation}
where the values of $\dot \gamma_j$, $C_j$ and $n_j$  are taken in a way to preserve continuity of $\eta_\text{a}$ and provide the best approximation of  $\eta_\text{a}^{(\text{g})}$ for the chosen value of $N$. Thus, the analyzed range of shear rates is subdivided into $N+1$ intervals ($N$ limiting values of the shear rate $\dot \gamma_j$) including:
\begin{itemize}
\item{two sections (for $|\dot \gamma |<|\dot \gamma_1|$ and $|\dot \gamma|>|\dot \gamma_N|$) where the Newtonian models hold with viscosities defined by $\eta_0$ and $\eta_\infty$, respectively; }
\item{$N-1$ sections where the power-law behavior is assumed.}
\end{itemize}
Naturally, for growing $N$ the quality of approximation is improved.

\begin{remark}
In order to have a direct analogy to the truncated power-law and Carreau examples we used the cut-off viscosities in the general relation \eqref{eta_ap}. Obviously, it is sufficient to employ only  the power-law type representation with setting $n_j=1$, if necessary. Alternatively, if the respective viscosity limit does not exist,  one can introduce in \eqref{eta_ap} $|\dot\gamma_1|=0$ or $|\dot\gamma_N|=\infty$ accordingly. For example in the case of Ellis model \cite{Bird_1987} the latter value is used.
\end{remark}

With such a representation of the apparent viscosity up to $N+1$ shear rate layers can appear across each of the channel's symmetrical parts depending on the magnitudes of $w$ and $dp/dx$ (there were at most three such layers with the truncated power-law model). Thicknesses of these layers are defined as (respective values can be identified with those shown in Fig. \ref{channel} when assuming $N=2$):
\begin{itemize}
\item{for the Newtonian-type layer in the core of the flow
\begin{equation}
\label{delta_1_ap}
\delta_1=\left( \frac{dp}{dx}\right)^{-1}\eta_0 \dot \gamma_1,
\end{equation}}
\item{for the power-law layers in the range $|\dot \gamma_1|<|\dot \gamma|<|\dot \gamma_N|$
\begin{equation}
\label{delta_j_ap}
\delta_{j+1}=\left(-\frac{dp}{dx}\right)^{-1}C_j\left[(-\dot \gamma_{j+1})^{n_j} -(-\dot \gamma_{j})^{n_j}\right], \quad j=1,...,N-1,
\end{equation}}
\item{for the Newtonian layer adjacent to the channel wall
\begin{equation}
\label{delta_N_ap}
\delta_N=\frac{w}{2}-\sum_{j=1}^{N}\delta_j.
\end{equation}}
\end{itemize}
Naturally, just as it was in the case of the truncated power-law, some of these layers may not be present or some of them can be reduced by the overall height of the channel. 

Now, having defined the apparent viscosity by \eqref{eta_ap}, we can analytically compute the velocity profiles over each of the component layers. The respective formulae are:
\begin{itemize}
\item{the Newtonian layer in the core of the flow
\begin{equation}
\label{v_1}
v^{(1)}(y)=v_1-\frac{1}{2\eta_0}\frac{dp}{dx}\left(\delta_1^2-y^2 \right), \quad y \in[0,\delta_1],
\end{equation}}
\item{the power-law layers in the range $|\dot \gamma_1|<|\dot \gamma|<|\dot \gamma_N|$
\begin{equation}
\label{v_j}
v^{(j+1)}(y)=v_{j+1}-\frac{n_j}{n_j+1}\left(\frac{dp}{dx}\right)^{-1}C_j\left\{\left[(-\dot \gamma_j)^{n_j}-\frac{1}{C_j}\frac{dp}{dx}\delta_{j+1} \right] ^\frac{n_j+1}{n_j}-\left[(-\dot \gamma_j)^{n_j}-\frac{1}{C_j}\frac{dp}{dx}(y-y_j) \right] ^\frac{n_j+1}{n_j}\right\},
\end{equation}
\[
y \in \left[y_j,y_{j+1}\right], \quad  j=1,...,N-1,
\]}
\item{the Newtonian layer adjacent to the channel wall
\begin{equation}
\label{v_N+1}
v^{(N+1)}(y)=-\dot \gamma_N\left(\frac{w}{2}-y\right)-\frac{1}{2\eta_\infty}\frac{dp}{dx}\left[\left(\frac{w}{2}-y_N \right)^2-\left(y-y_N \right)^2\right], \quad y \in \left[y_N, \frac{w}{2}\right],
\end{equation}}
\end{itemize}
where:
\[
y_j=\sum_{k=1}^j\delta_k.
\]
As was the case in the truncated power-law problem, the interfacial values of the velocity are determined as:
\begin{equation}
\label{v_inter}
v_j=v^{(j+1)}\left(y_j\right), \quad j=1,...,N-1.
\end{equation}
Note that both the velocity obtained is this way and its derivative are continuous over the channel height.

Having the above approximation of the velocity profile one can analytically compute the average fluid flow rate through the channel cross section according to formula \eqref{Q_def} which yields:
\begin{equation}
\label{Q_Carreau}
Q=2\left[ \int_0^{y_1} v^{(1)}dy +\sum_{j=1}^{N-1}\int_{y_j}^{y_{j+1}}v^{(j+1)}dy+\int_{y_N}^{w/2} v^{(N+1)}dy\right],
\end{equation}
where the respective integrals are given by the following relations:
\begin{equation}
\label{Q_11}
 \int_0^{y_1} v^{(1)}dy=v_1\delta_1-\frac{1}{3\eta_0}\frac{dp}{dx}\delta_1^3,
\end{equation}

\begin{equation}
\label{Q_j}
\begin{aligned}
\int_{y_j}^{y_{j+1}}v^{(j+1)}dy={} & v_{j+1}\delta_{j+1}-\frac{n_j}{n_j+1}C_j\left(\frac{dp}{dx}\right)^{-1}\Bigg\{\left(\left(-\dot\gamma_j\right)^{n_j}-\frac{1}{C_j}\frac{dp}{dx}\delta_{j+1} \right)^\frac{1+n_{j+1}}{n_j}\delta_{j+1}   \\
      & +\frac{n_j}{2n_j+1}C_j\left(\frac{dp}{dx} \right)^{-1} \Bigg[ \left(\left( -\dot \gamma_j\right)^{n_j}-\frac{1}{C_j}\frac{dp}{dx}\delta_{j+1} \right)^\frac{2n_j+1}{n_j} -\left(-\dot \gamma_j \right)^{2n_j+1}\Bigg]\Bigg\}.
\end{aligned}
\end{equation}

\begin{equation}
\label{Q_N}
\int_{y_N}^{w/2} v^{(N+1)}dy=-\frac{1}{2}\dot\gamma_N\delta_{N+1}^2-\frac{1}{3\eta_\infty}\frac{dp}{dx}\delta_{N+1}^3,
\end{equation}
In this way, computation of the average fluid flow rate through the channel is reduced to summation of up to $N+1$ terms of the form \eqref{Q_11}-\eqref{Q_j}.

Clearly, for the predefined values of $w$ and $dp/dx$ the number of the shear rate layers can be smaller than $N+1$. In such a case respective formulae for the fluid velocity and layers' thicknesses should be adjusted accordingly in the same manner as that shown for the truncated power-law model. Consequently, the number of component integrals in the relation \eqref{Q_Carreau} can be lower than $N+1$.

By using the above simple analytical relations one can obtain an approximate solution for the generalized Newtonian fluid. The accuracy of solution is limited only by the accuracy to which the approximated viscosity \eqref{eta_ap} mimics the general viscosity $\eta_\text{a}^{(\text{g})}$. All other computations amount to analytical transformations and as such do not introduce any error. The algorithm of solution includes the following steps:
\begin{enumerate}
\item{The preconditioning step. Here, for a predefined value of $N$, approximation \eqref{eta_ap} is constructed. As a result one obtains $N-1$ coefficients $C_j$ and corresponding $n_j$ for $N$ limiting values of $\dot\gamma_j$.}
\item{For given values of $w$ and $dp/dx$ and approximation of the apparent viscosity \eqref{eta_ap}, the number and sizes of respective shear rate layers are determined  according to formulae \eqref{delta_1_ap}--\eqref{delta_N_ap}. }
\item{The velocity profile is reconstructed from relations \eqref{v_1}--\eqref{v_N+1}. Naturally, if one is interested only in the value of the average fluid flow rate, $Q$, this step can be omitted.}
\item{The fluid flow rate, $Q$, is computed according to \eqref{Q_Carreau}--\eqref{Q_j}. This step amounts to algebraic summation of up to $N+1$ terms of the type  \eqref{Q_11}--\eqref{Q_j}.}
\end{enumerate}
Note that the whole scheme is based on simple analytical transformations and its numerical implementation is straightforward while the computational cost is very low (no need for iterative computations or solving systems of equations etc.). Moreover, the algorithm is not restricted to any particular rheological model and can be used even with experimental data. 

\section{Numerical results}
\label{num_res}

In this section we analyze the accuracy and efficiency of computations achievable with the algorithm defined in Section \ref{apr_sec}. The analysis will be performed using the example of the Carreau fluid. The Carreau law parameters are defined in Table \ref{par_table} and the channel height is $w=10^{-3} \ \text{m}$. As a reference solution for comparison we use the numerical solution obtained by the scheme presented in \ref{ap_num_sol} for spatial domain discretized by 200 points ($M=200$). According to the analysis performed in  \ref{ap_num_sol}, it yields the maximal relative errors lower than $10^{-10}$ for both, the fluid velocity, $v$, and the fluid flow rate, $Q$. All the computations are carried out in the Matlab environment. 

\subsection{Approximation of the apparent viscosity}

Constructing the approximation \eqref{eta_ap} is considered  a preconditioning step in the proposed algorithm. The Carreau apparent viscosity, $\eta_\text{a}^{(\text{c})}$, is defined by the relation \eqref{carreau_def}. Generally, finding the optimal representation for mimicking the Carreau law can be quite a complicated problem. In this paper we employ the following simple approach. First, for a predefined value of $N$ we determine the limiting shear rates $\dot \gamma_1$ and $\dot \gamma_N$ by assuming that the relative deviations between $\eta_\text{a}$ (i.e. the approximation \eqref{eta_ap}) and  $\eta_\text{a}^{(\text{c})}$:
\[
\delta \eta_\text{a}=\left|\frac{\eta_\text{a}-\eta_{\text{a}}^{(\text{c})}}{\eta_{\text{a}}^{(\text{c})}}\right|
\]
have some prescribed value $\varepsilon$ for these limiting rates, i. e. $\delta \eta_\text{a}(\dot \gamma_1)=\delta \eta_\text{a}(\dot \gamma_N)=\varepsilon$. The values of $\dot \gamma_j$ are specified as:
\[
\dot \gamma_j=\dot \gamma_{j-1}\left( \frac{\dot \gamma_N}{\dot \gamma_1}\right)^{1/N},
\]
which results in their uniform distribution on the logarithmic scale. Over each section $|\dot \gamma_j|<|\dot \gamma|<|\dot \gamma_{j+1}|$ respective values of $C_j$ and $n_j$ are found from the conditions:
\begin{equation}
\label{gam_cont}
C_j \left(-\dot \gamma_j\right)^{n_j-1}=C_{j-1} \left(-\dot \gamma_j\right)^{n_{j-1}-1},
\end{equation}
\begin{equation}
\label{gam_int}
\int_{\dot \gamma_j}^{\dot \gamma_{j+1}} \eta_\text{a}d\dot\gamma=\int_{\dot \gamma_j}^{\dot \gamma_{j+1}} \eta_\text{a}^{(\text{c})}d\dot\gamma.
\end{equation}
Note that \eqref{gam_cont} provides continuity of $\eta_\text{a}$.

\begin{figure}[htb!]
\begin{center}
\includegraphics[scale=0.38]{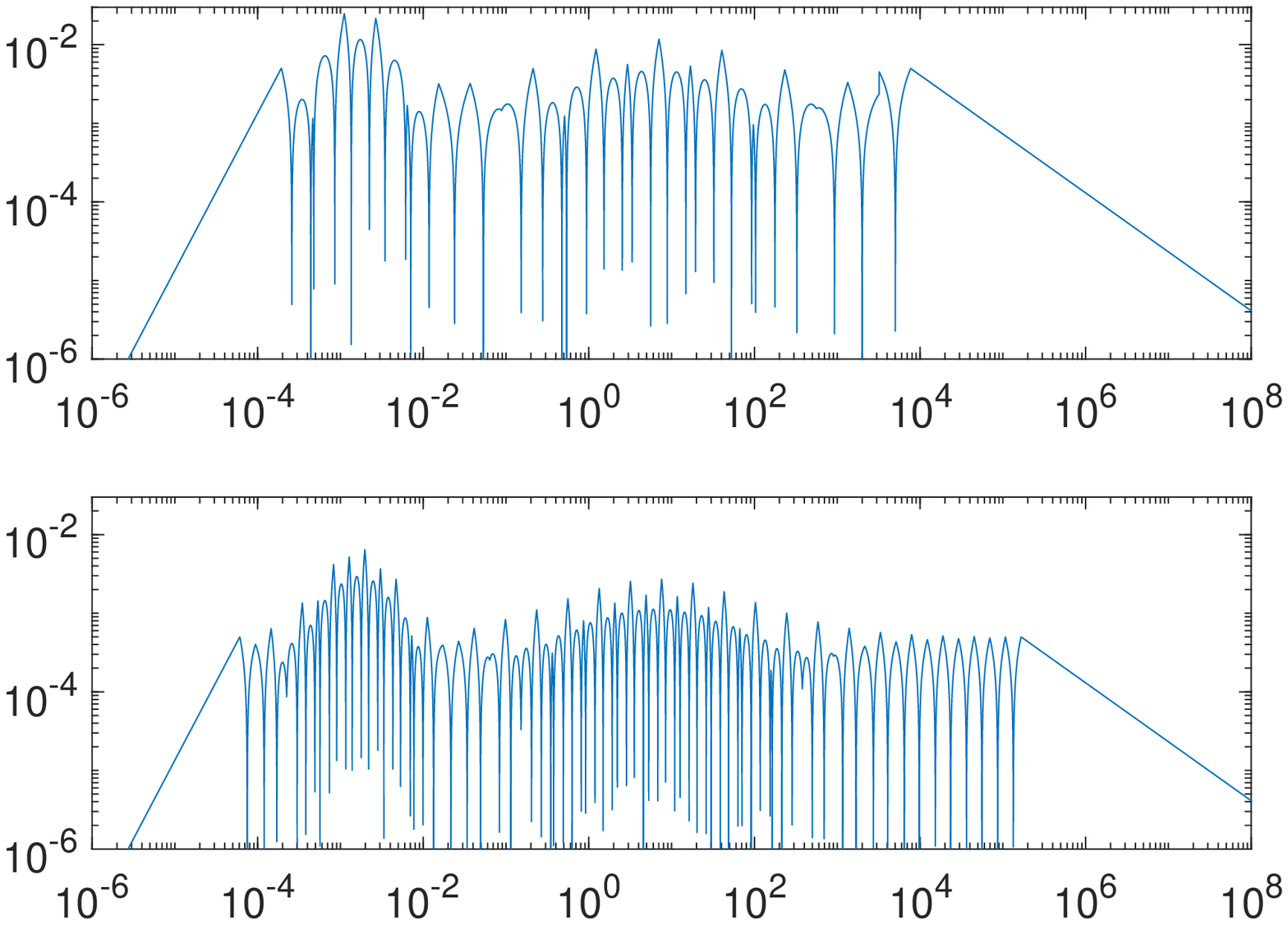}
\hspace{0mm}
\includegraphics[scale=0.38]{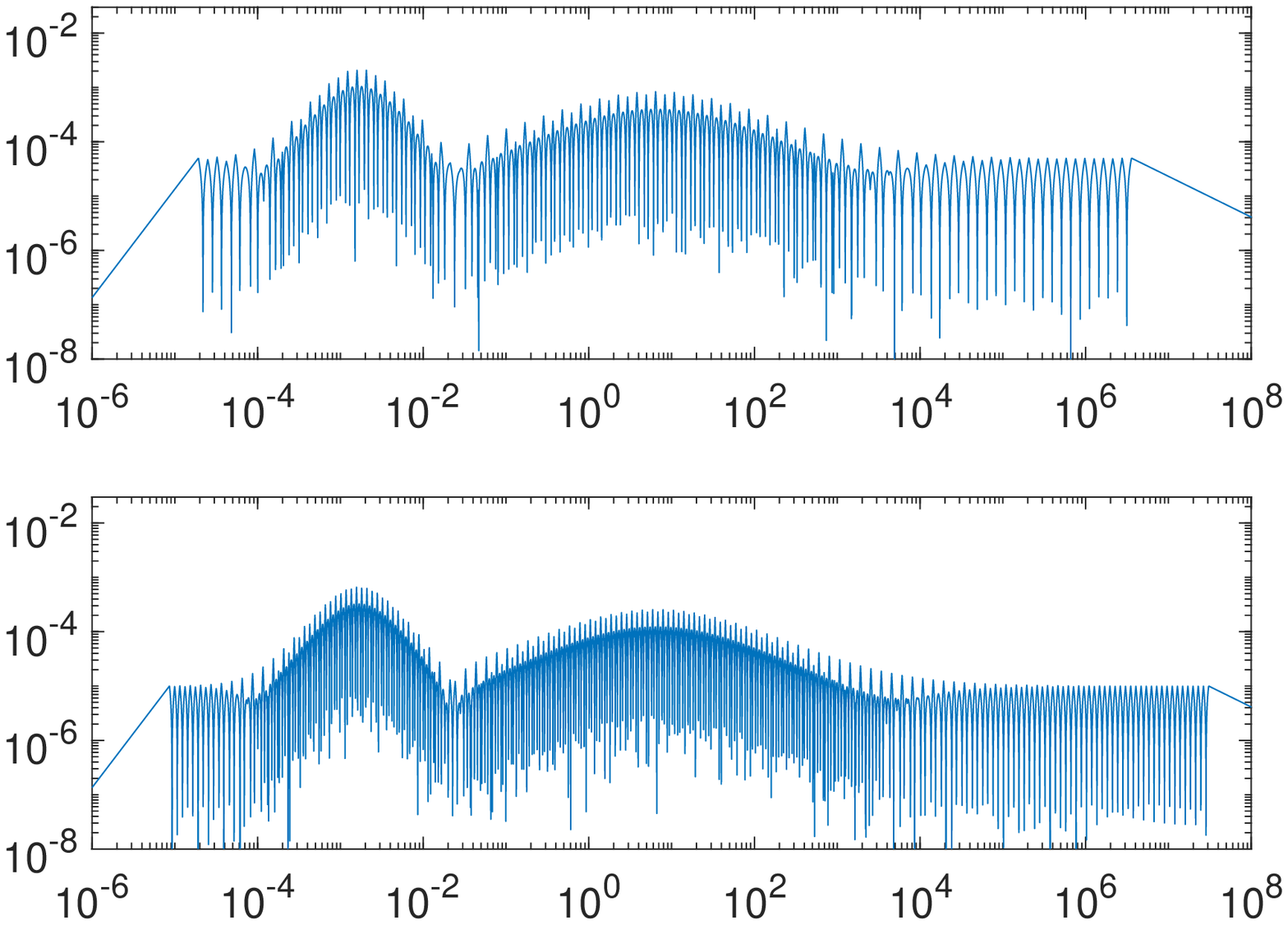}
\put(-338,-5){$|\dot \gamma|$}
\put(-110,-5){$|\dot \gamma|$}
\put(-446,135){$\textbf{a)}$}
\put(-446,60){$\textbf{b)}$}
\put(-220,135){$\textbf{c)}$}
\put(-220,60){$\textbf{d)}$}
\put(-285,127){$N=20$}
\put(-285,55){$N=50$}
\put(-65,55){$N=200$}
\put(-65,127){$N=100$}
\put(-446,109){$\delta \eta_\text{a}$}
\put(-220,110){$\delta \eta_\text{a}$}
\put(-446,36){$\delta \eta_\text{a}$}
\put(-220,36){$\delta \eta_\text{a}$}
\caption{The quality of approximation of the apparent viscosity, $\delta \eta_\text{a}$, according to formula \eqref{eta_ap} .}
\label{Carreau_apr}
\end{center}
\end{figure}

In Fig. \ref{Carreau_apr} we present the accuracy of approximation of the Carreau law by the relation \eqref{eta_ap} for four values of $N=\{20,50,100,200\}$. As can be seen, the average error  decreases with growing $N$. However, even for the same $N$ the local accuracy  can differ by a few orders of magnitude. In Table \ref{eta_a_acc} we collate the maximal and average errors over the span $|\dot \gamma_1|<|\dot \gamma|<|\dot \gamma_N|$ , where the average errors are computed as:
\[
\hat{\delta \eta_\text{a}}=\frac{\int_{\dot \gamma_1}^{\dot \gamma_{N}} \left|\eta_\text{a}-\eta_\text{a}^{(\text{c})}\right|d\dot\gamma}{\int_{\dot \gamma_1}^{\dot \gamma_{N}} \eta_\text{a}^{(\text{c})}d\dot\gamma}.
\]
The results indicate that  adaptive meshing of the shear rate interval could improve the accuracy. 

\begin{table}
\begin{center}
\begin{tabular}{ |p{2cm}||p{2cm}|p{2cm}|  p{2cm}| p{2cm}| }
 \hline
$N$& 20 &50&100&200\\
 \hline
$\max \left(\delta \eta_\text{a}\right)$   & $2.48\cdot 10^{-2}$    &$6.43\cdot 10^{-3}$ &$2.07\cdot 10^{-3}$ &$6.56\cdot 10^{-4}$\\
 \hline
$\hat{\delta \eta_\text{a}}$ & $2.02\cdot 10^{-3}$    &$2.48\cdot 10^{-4}$ &$2.46\cdot 10^{-5}$ &$4.97\cdot 10^{-6}$\\
 \hline
\end{tabular}
\end{center}
\caption{The maximal and average errors of approximation of the Carreau law by the relation \eqref{eta_ap}.}
\label{eta_a_acc}
\end{table}

\subsection{The accuracy analysis}

\begin{figure}[htb!]
\begin{center}
\includegraphics[scale=0.38]{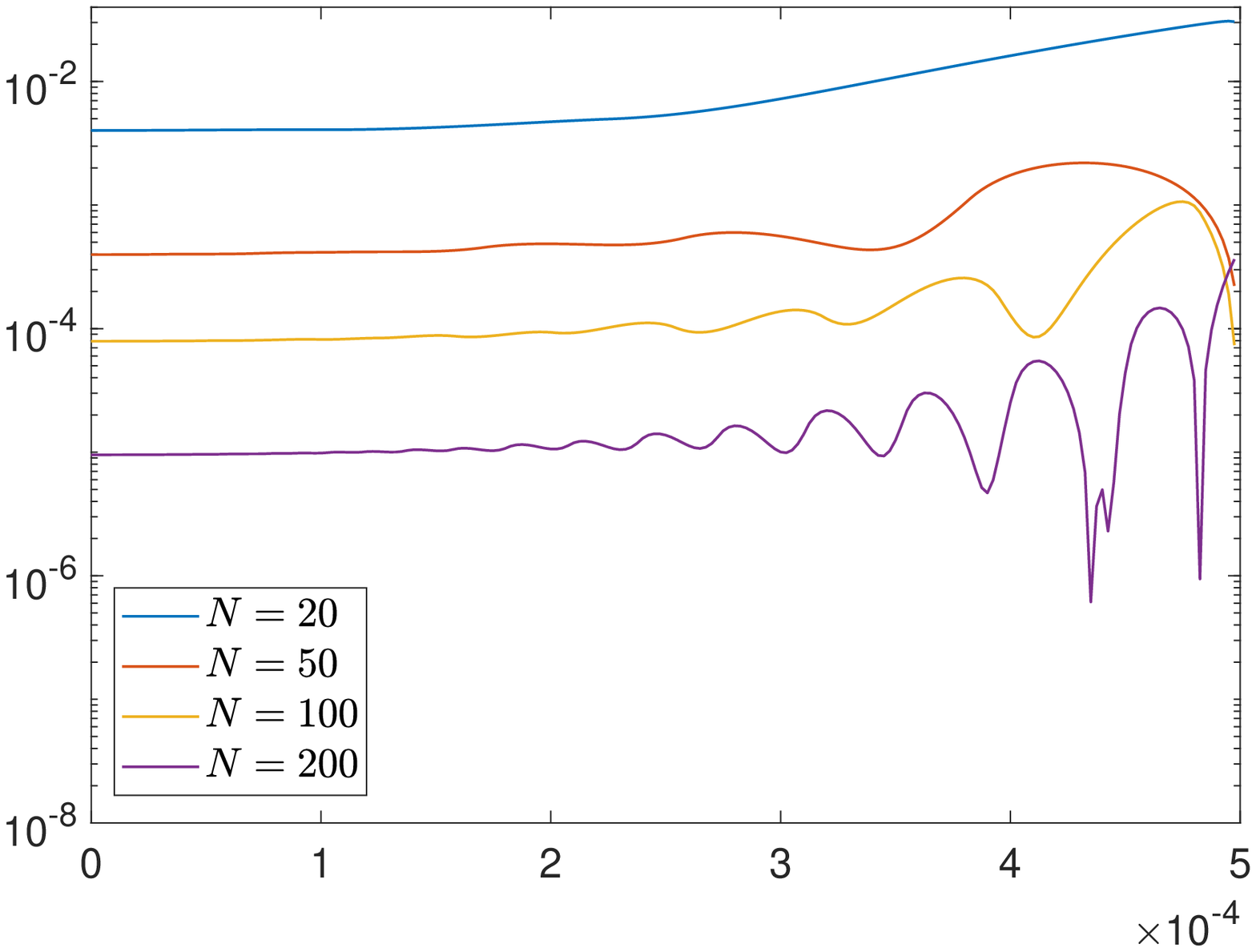}
\hspace{0mm}
\includegraphics[scale=0.38]{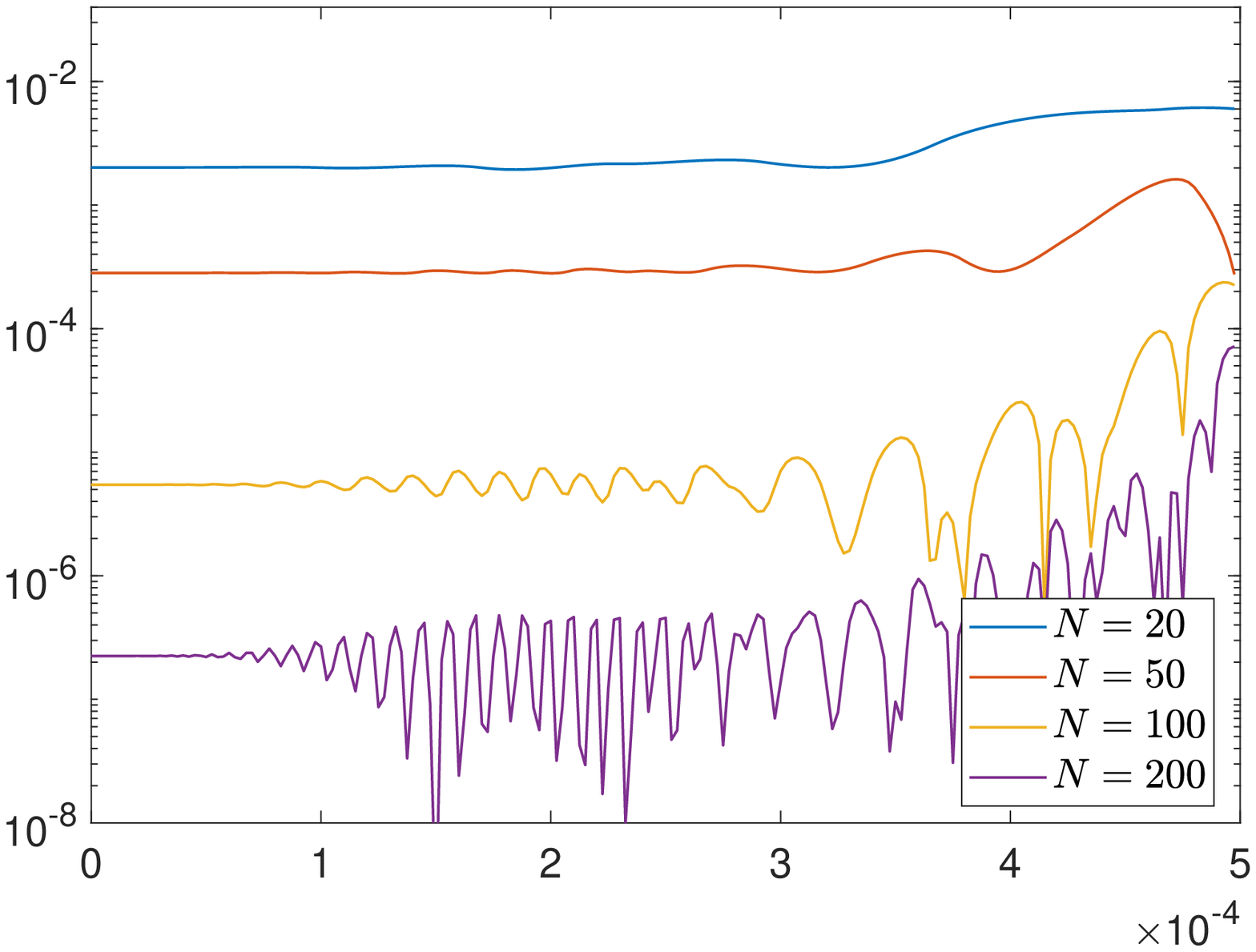}
\put(-338,-5){$y$}
\put(-110,-5){$y$}
\put(-446,135){$\textbf{a)}$}
\put(-220,135){$\textbf{b)}$}
\put(-446,75){$\delta v$}
\put(-220,75){$\delta v$}
\put(-385,130){$dp/dx=-1 \ \text{Pa/m}$}
\put(-150,130){$dp/dx=-5 \ \text{Pa/m}$}
\caption{The relative error of fluid velocity, $\delta v$, for: a) $dp/dx=-1 \ \text{Pa/m}$, b) $dp/dx=-5 \ \text{Pa/m}$.}
\label{del_v_1}
\end{center}
\end{figure}

\begin{figure}[htb!]
\begin{center}
\includegraphics[scale=0.38]{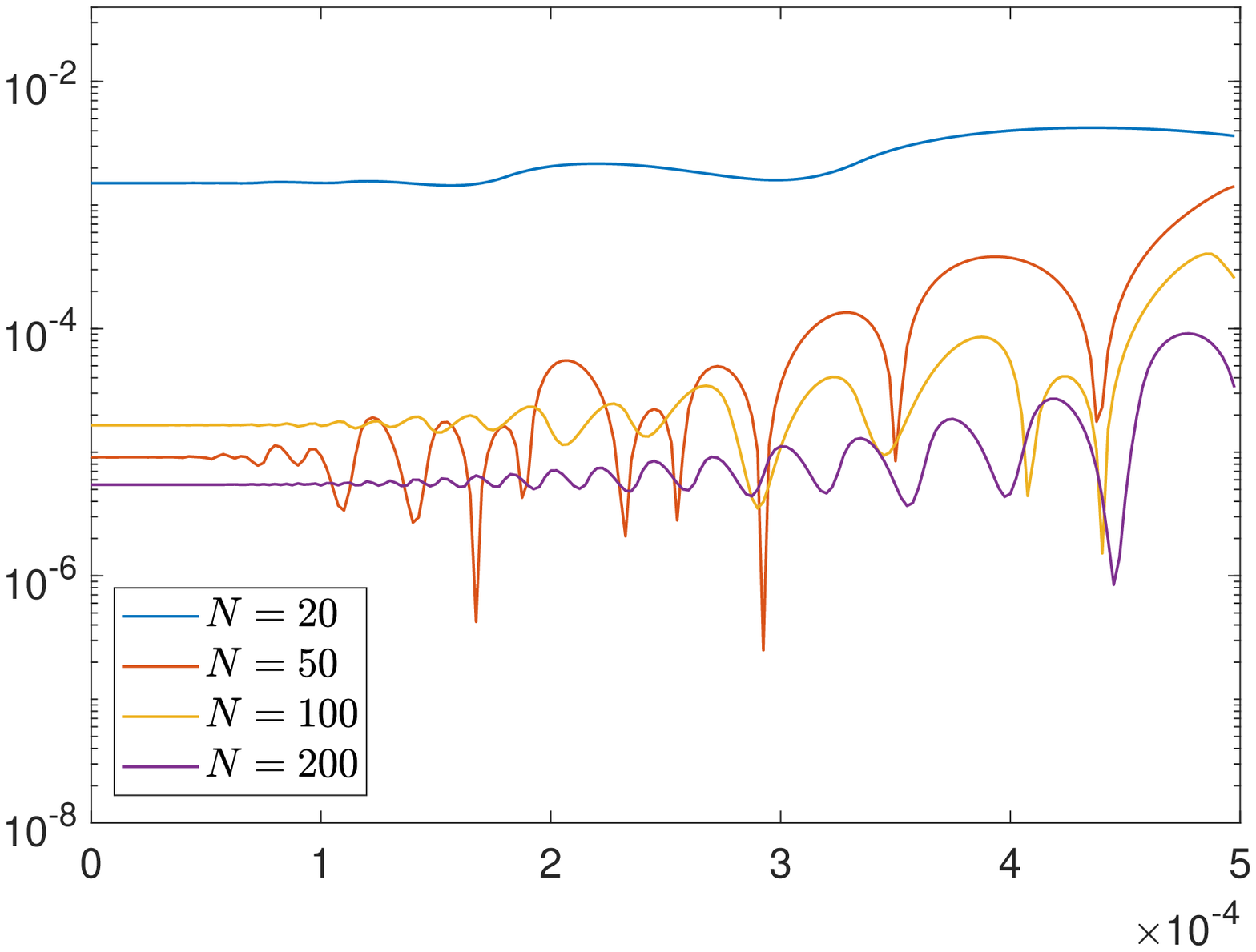}
\hspace{0mm}
\includegraphics[scale=0.38]{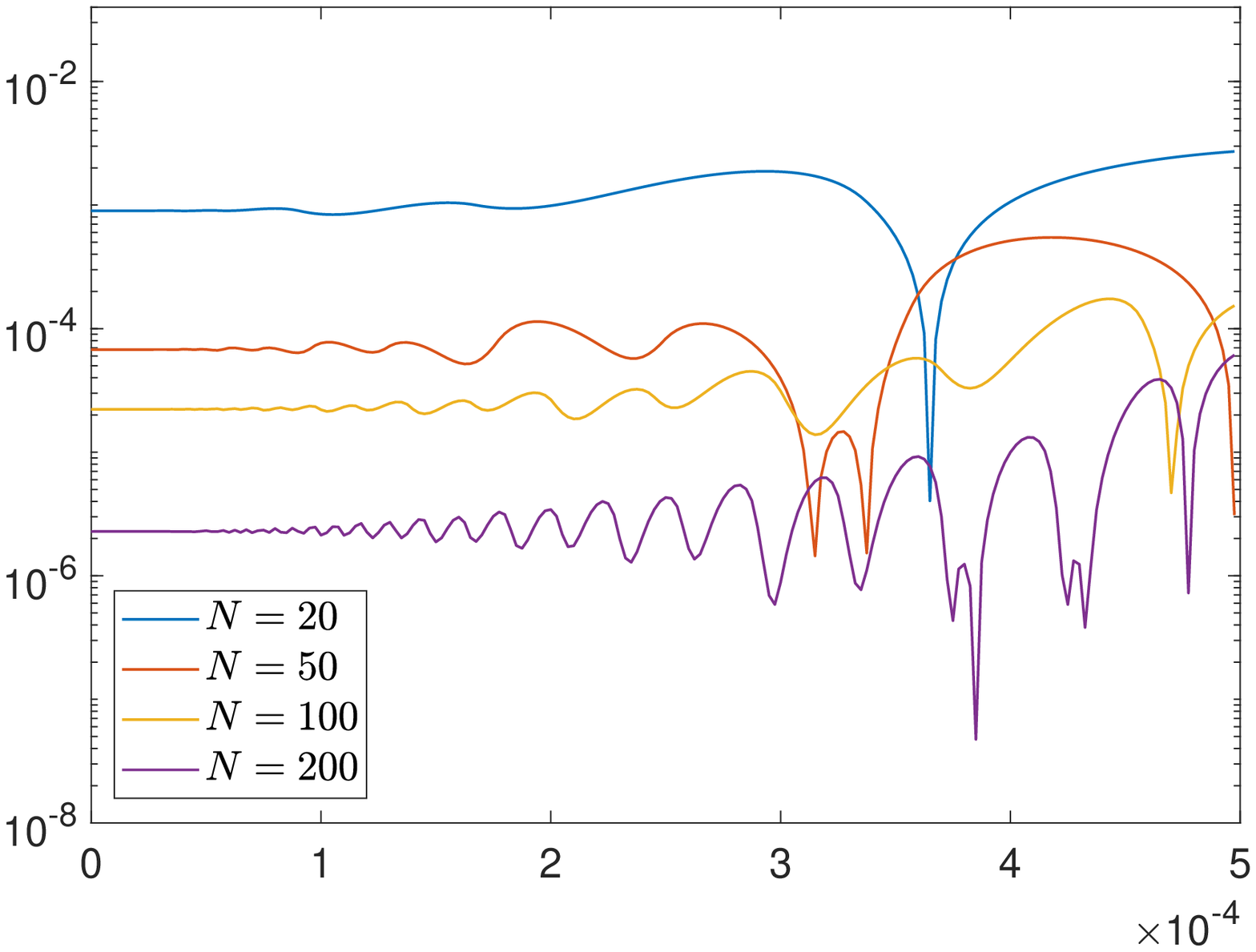}
\put(-338,-5){$y$}
\put(-110,-5){$y$}
\put(-446,135){$\textbf{a)}$}
\put(-220,135){$\textbf{b)}$}
\put(-446,75){$\delta v$}
\put(-220,75){$\delta v$}
\put(-375,127){$dp/dx=-75 \ \text{Pa/m}$}
\put(-150,127){$dp/dx=-150 \ \text{Pa/m}$}
\caption{The relative error of fluid velocity, $\delta v$, for: a) $dp/dx=-75 \ \text{Pa/m}$, b) $dp/dx=-150 \ \text{Pa/m}$.}
\label{del_v_2}
\end{center}
\end{figure}

Let us first analyze the accuracy of the fluid velocity. In Figs. \ref{del_v_1}-\ref{del_v_2} the relative errors of fluid velocity, $\delta v$, over the channel height  are depicted. The computations were performed for four values of the pressure gradient ranging from $-150 \ \text{Pa/m}$ to $-1 \ \text{Pa/m}$ . The approximation of the Carreau law in the form \eqref{eta_ap} is implemented for four variants of $N=\{20,50,100,200\}$.

It can be noted that, even for the simplest approximation of the apparent viscosity ($N=20$), the results are accurate. The maximal relative error of velocity of $3\%$ is obtained here for $dp/dx=-1 \ \text{Pa/m}$ in the proximity of the channel wall (note that $v\to0$ as $y\to w/2$). This value is only sightly higher than the maximal error of approximation of $\eta_\text{a}^{(\text{c})}$ that is $2.48\cdot 10^{-2}$ in this case (see Table \ref{eta_a_acc}). With growing magnitude of the pressure gradient the accuracy increases and the maximal error is well below $1\%$.  Naturally, this trend results from improving the quality of approximation \eqref{eta_ap} for increasing values of the shear rate (which is depicted in Fig. \ref{Carreau_apr}a)). For the higher order approximations of the apparent viscosity ($N>20$) the maximal errors are much lower than $1\%$ regardless of the value of $N$ and $dp/dx$. For example, even with $N=50$ the maximal error is of the order $10^{-3}$ while the  average errors are of one order of magnitude lower. The maximal and average relative errors of velocity are collated in Table \ref{v_ac_errors}. The average error is computed as:
\[
\delta \hat v=\frac{\int_0^{w/2}\left|v-v_\text{b} \right|dy}{\int_0^{w/2}v_\text{b}dy},
\]
where $v_\text{b}$ is the benchmark  velocity.

\begin{table}[htb!]
\begin{center}
\begin{tabular}{cl|l|l|l|l|}
\cline{3-6}
\multicolumn{1}{l}{}                       &     & \multicolumn{4}{c|}{$|dp/dx|, \ \text{Pa/m}$}                                                                       \\ \cline{3-6} 
\multicolumn{1}{l}{}                       &     & \multicolumn{1}{c|}{1} & \multicolumn{1}{c|}{5} & \multicolumn{1}{c|}{75} & \multicolumn{1}{c|}{150} \\ \cline{3-6} 
\multicolumn{1}{l}{}                       &     & \multicolumn{4}{c|}{$\max\left(\delta v\right)$}                                                     \\ \hline
\multicolumn{1}{|c|}{\multirow{4}{*}{$N$}} & 20  & $3.08\cdot10^{-2}$     & $6.14\cdot10^{-3}$     & $4.23\cdot10^{-3}$      & $2.72\cdot10^{-3}$       \\ \cline{2-6} 
\multicolumn{1}{|c|}{}                     & 50  & $2.19\cdot10^{-3}$     & $1.62\cdot10^{-3}$     & $1.42\cdot10^{-3}$      & $5.47\cdot10^{-4}$       \\ \cline{2-6} 
\multicolumn{1}{|c|}{}                     & 100 & $1.07\cdot10^{-3}$     & $2.37\cdot10^{-4}$     & $4.03\cdot10^{-4}$      & $1.74\cdot10^{-4}$       \\ \cline{2-6} 
\multicolumn{1}{|c|}{}                     & 200 & $3.65\cdot10^{-4}$     & $7.18\cdot10^{-5}$     & $9.12\cdot10^{-5}$      & $6.12\cdot10^{-5}$       \\ \hline
\multicolumn{1}{l}{}                       &     & \multicolumn{4}{c|}{$\delta \hat v$}                                                                 \\ \hline
\multicolumn{1}{|c|}{\multirow{4}{*}{$N$}} & 20  & $6.47\cdot10^{-3}$     & $2.51\cdot10^{-3}$     & $1.97\cdot10^{-3}$      & $1.11\cdot10^{-3}$       \\ \cline{2-6} 
\multicolumn{1}{|c|}{}                     & 50  & $5.71\cdot10^{-4}$     & $3.43\cdot10^{-4}$     & $5.88\cdot10^{-5}$      & $1.10\cdot10^{-4}$       \\ \cline{2-6} 
\multicolumn{1}{|c|}{}                     & 100 & $1.17\cdot10^{-4}$     & $8.49\cdot10^{-6}$     & $2.46\cdot10^{-5}$      & $3.17\cdot10^{-5}$       \\ \cline{2-6} 
\multicolumn{1}{|c|}{}                     & 200 & $1.40\cdot10^{-5}$     & $5.29\cdot10^{-7}$     & $7.71\cdot10^{-6}$      & $3.39\cdot10^{-6}$       \\ \hline
\end{tabular}
\end{center}
\caption{The maximal, $\max \left(\delta v\right)$, and average, $\delta \hat v$, relative errors of the fluid velocity.}
\label{v_ac_errors}
\end{table}

\begin{figure}[htb!]
\begin{center}
\includegraphics[scale=0.38]{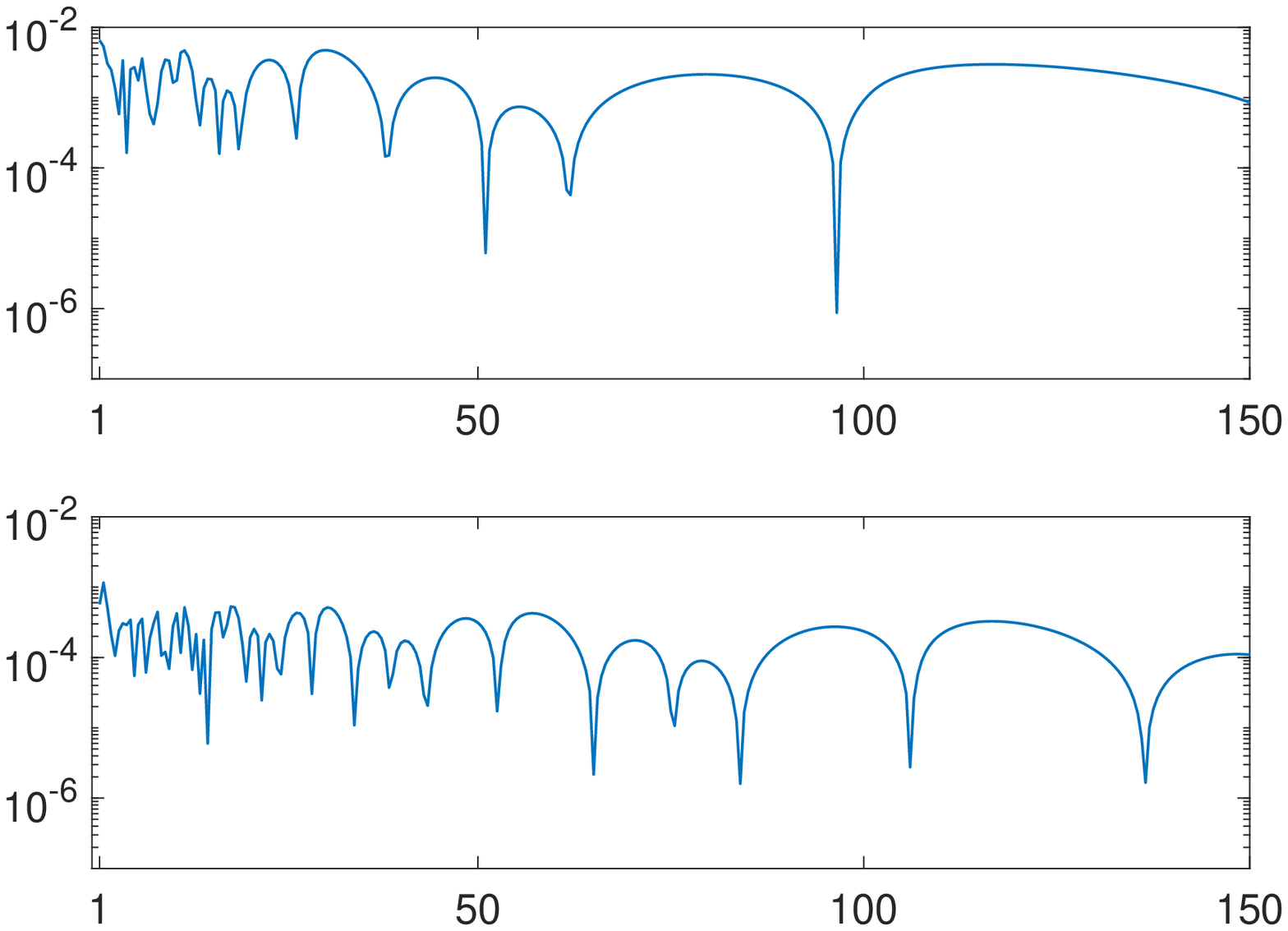}
\hspace{0mm}
\includegraphics[scale=0.38]{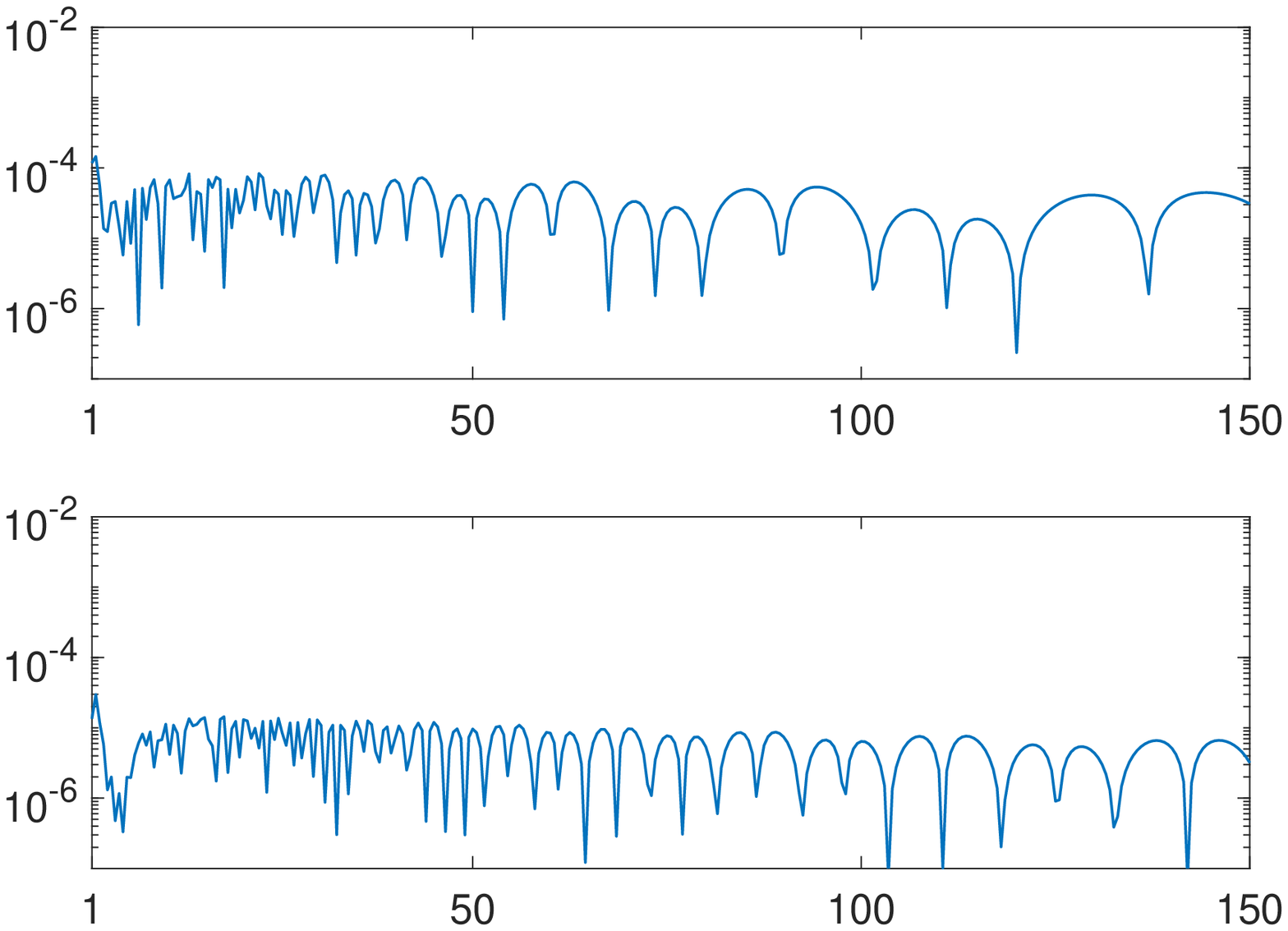}
\put(-338,-5){$|dp/dx|$}
\put(-110,-5){$|dp/dx|$}
\put(-446,135){$\textbf{a)}$}
\put(-446,60){$\textbf{b)}$}
\put(-220,135){$\textbf{c)}$}
\put(-220,60){$\textbf{d)}$}
\put(-446,109){$\delta Q$}
\put(-220,110){$\delta Q$}
\put(-446,36){$\delta Q$}
\put(-220,36){$\delta Q$}
\put(-290,120){$N=20$}
\put(-290,55){$N=50$}
\put(-70,55){$N=200$}
\put(-70,127){$N=100$}
\caption{The relative error of the fluid flow rate, $\delta Q$.}
\label{del_q_1}
\end{center}
\end{figure}

The relative errors of the fluid flow rate, $\delta Q$, for the pressure gradient ranging from $-150 \ \text{Pa/m}$ to $-1 \ \text{Pa/m}$  are depicted in Fig. \ref{del_q_1}. In this case even for $N=20$ the maximal error is much lower than $1\%$. With growing $N$ the accuracy is constantly improved. The average errors computed as (the subscript `$\text{b}$' refers to the benchmark value, i.e. the one obtained with the method from  \ref{ap_num_sol}):
\[
\delta \hat Q=\frac{\int_{-150}^{-1}\left|Q(p')-Q_\text{b}(p')\right|dp' }{\int_{-150}^{-1}Q_\text{b}(p')dp'}
\]
are reduced from $1.80\cdot 10^{-3}$ for $N=20$ to $4.97\cdot 10^{-6}$ for $N=200$. The errors' magnitudes are collected in Table \ref{Q_err}.

\begin{table}[]
\begin{center}
\begin{tabular}{|l|l|l|l|l|}
\hline
$N$                         & 20                  & 50                  & 100                 & 200                 \\ \hline
$\max\left(\delta Q\right)$ & $6.51\cdot 10^{-3}$ & $5.71\cdot 10^{-4}$ & $1.17\cdot 10^{-4}$ & $1.44\cdot 10^{-5}$ \\ \hline
$\delta \hat Q$             & $1.80\cdot 10^{-3}$ & $1.62\cdot 10^{-4}$ & $2.81\cdot 10^{-5}$ & $4.94\cdot 10^{-6}$ \\ \hline
\end{tabular}
\end{center}
\caption{The maximal and average errors of the fluid flow rate.}
\label{Q_err}
\end{table}

\subsection{The efficiency of computations}

The above analysis proved that even with the simplest considered  accuracy of approximation of the apparent viscosity \eqref{eta_ap} ($N=20$) one obtains the accuracy of solution sufficient for most practical applications. This accuracy can be easily improved by increasing $N$. Let us now compare the efficiency of computations with other available methods. The following schemes are used for this comparison: 
\begin{itemize}
\item{The semi-analytical solution for the fluid flow rate obtained according to the algorithm given in \cite{Sochi_2015}. The solution utilizes the hypergeometric Gauss functions whose generation in the Matlab environment is rather time consuming. In our implementation the computations of these functions by means of the standard Matlab subroutines took $99.7\%$ of the overall computational time (i.e. only $0.3\%$ of the computational time was expended for other operations).  }
\item{The numerical solution computed according to the scheme described in \ref{ap_num_sol}. Here the full velocity profile is recreated with a subsequent integration to obtain the fluid flow rate.  The computations are carried out with $M=200$ nodal points across the spatial domain.}
\item{The new approximate solution introduced in Section \ref{apr_sec} for four variants of $N$ (as previously $N=\{20,50,100,200\}$).}
\end{itemize}
The criterion used to compare the efficiency of computations is the time of computing the fluid flow rate, $Q$, for an imposed pressure gradient. The computations were carried out for $-150 \ \text{Pa/m}\leq dp/dx \leq -1 \ \text{Pa/m}$ with a resolution of $0.5 \ \text{Pa/m}$. For each of the listed methods an arithmetic mean of 300 computational times was taken.

 Let us introduce the following normalized times:
\[
\tau_1=\frac{t_1}{t_3}, \quad \tau_2=\frac{t_2}{t_3},
\]
where $t_1$ is the average computational time for the method provided in \cite{Sochi_2015}, $t_2$ is the average computational time for the method from \ref{ap_num_sol}, whereas $t_3$ denotes the average time expended for the computations according to the algorithm introduced in Section \ref{apr_sec}. In this way $\tau_1$ and $\tau_2$ inform us to what degree the respective scheme (the first and the second one) is more or less efficient than the newly introduced algorithm. The values of the normalized times are listed in Table \ref{tau_tab}.

\begin{table}
\begin{center}
\begin{tabular}{ |p{0.5cm}||p{1cm}|p{1cm}|  p{1cm}| p{1cm}| }
 \hline
$N$& 20 &50&100&200\\
 \hline
$\tau_1$   &774    &676 &476 &310\\
 \hline
$\tau_2$ & 323    &290 &204 &133\\
 \hline
\end{tabular}
\end{center}
\caption{The normalized times of computations for: $\tau_1$ - semi-analytical solution from \cite{Sochi_2015}, $\tau_2$ - the numerical solution from \ref{ap_num_sol}.}
\label{tau_tab}
\end{table}

The obtained results prove that the newly introduced scheme has an immense advantage in efficiency over the other two. Indeed, even when using the high order approximation of the apparent viscosity  ($N=200$), it takes 133 less time to compute the solution than in the case of the algorithm from  \ref{ap_num_sol}. The respective factor for the semi-analytical solution from \cite{Sochi_2015} is even greater and amounts to 310. Let us reiterate that the approximate solution obtained in this way is very accurate and according to the data presented in Tables \ref{v_ac_errors}-\ref{Q_err} yields the errors of the level of $10^{-6}$ for both, the fluid velocity and the fluid flow rate. Naturally, when decreasing $N$ the advantage of the new method becomes even greater, giving for $N=20$ over 300 times higher efficiency that the scheme of \ref{ap_num_sol} and over 700 higher than the one from \cite{Sochi_2015}. These figures are achieved with the accuracy of solution still sufficient for most practical applications. 

\section{Final conclusions}
\label{disc}

In this paper a problem of a generalized Newtonian fluid flow in a flat channel was analyzed. A new method of solution was introduced which assumes  continuous approximation of the apparent viscosity by power-law functions with subsequent analytical integration to obtain the velocity profile and the fluid flow rate. The accuracy and efficiency of computations were investigated using an example of the Carreau fluid. 

The following conclusions can be drawn from the conducted analysis:
\begin{itemize}
\item{The proposed method constitutes a highly efficient and flexible tool for solving the problem of  a generalized Newtonian fluid flow.}
\item{The accuracy of obtained solution depends only on the quality of approximation of the apparent viscosity characteristics. Even for the simplest considered variant of the latter the overall accuracy is sufficient for most  practical applications. It can be increased to any desired level by using higher order approximations of the apparent viscosity.  }
\item{Although the proposed method of solution was demonstrated for the example of Carreau fluid, it can be easily adopted to other rheological characteristics including those given only by discrete data (e.g. experimental characteristics). }
\item{The new method of solution proved to be highly efficient. It requires a few orders of magnitude lower time of computations than a semi-analytical solution used for comparison. The method does not necessitate evaluation of special functions,  iterative computations or solving systems of algebraic equations.}
\item{The method of solution can be directly adopted for other straight conduits of simple geometries, such as circular or elliptic channels.}
\item{Due to its efficiency, the new method is recommended for problems where multiple evaluations of the fluid flow rate or velocity are required (e.g. in the hydraulic fracture problems).}
\item{The proposed procedure can be successfully used as a numerical substitute for the Poiseulle-type formulae in problems where no analytical solution exists.}
\end{itemize}

\section*{Acknowledgments}
\noindent
The author is thankful to Prof. Panos Papanastasiou, Prof. Gennady Mishuris and Dr. Monika Perkowska for their useful comments and discussions.

\vspace{5mm}
\noindent
{\bf Funding:} 
This work was funded by European Regional Development Fund and the Republic of Cyprus
through the Research Promotion Foundation (RESTART 2016 - 2020 PROGRAMMES, Excellence Hubs,
Project EXCELLENCE/1216/0481).

\appendix

\section{Numerical solution to the problem of Carreau fluid flow}
\label{ap_num_sol}

The numerical solution to the problem with the Carreau law is obtained in the following way. Just as was done in the main body of the paper, we consider only one symmetrical part of the channel $y \in [0,w/2]$. We discretize it by $M$ uniformly spaced nodal points $y_i$ with $i=1,...,M$.

When formally integrating equation \eqref{gov_ode} with respect to $y$ over the span $[0,y]$ under the boundary condition \eqref{BCs}$_1$ one arrives at the relation:
\begin{equation}
\label{ode_int}
\eta_a\left(\dot \gamma\right)\dot \gamma=\frac{dp}{dx}y.
\end{equation}
On the left hand side of \eqref{ode_int} the definition \eqref{GNF} was used. After simple rearrangement of equation \eqref{ode_int} and another analytical integration from $y$ to $w/2$ with the boundary condition \eqref{BCs}$_2$ we obtain the following formula for the fluid velocity:
\begin{equation}
\label{v_form}
v(y)=-\frac{dp}{dx}\int_y^{w/2}\frac{y}{\eta_\text{a}(\dot \gamma)}dy.
\end{equation}

For every point of the spatial domain, $y_i$, equation \eqref{ode_int} can be treated as an algebraic equation with respect to $\dot \gamma_i=\dot \gamma (y_i)$:
\begin{equation}
\label{gamma_dis}
\eta_\text{a}(\dot \gamma_i)\gamma_i=\frac{dp}{dx}y_i, \quad i=1,...,M.
\end{equation}
This equation is solved iteratively. The iterations are stopped when the relative difference between two consecutive values of $\dot \gamma$ falls below $10^{-12}$. After obtaining a discrete characteristics $\dot \gamma(y)$ it is used to approximate the integrand from equation \eqref{v_form} by cubic splines. Subsequent integration according to the formula yields the values of velocity at the respective spatial points $y_i$. The fluid flow rate, $Q$, is computed by integration of the velocity profile in line with formula \eqref{Q_def}.

We verify the accuracy of computations by the constructed scheme using two benchmark examples. The first one is an analytical solution obtained for the power-law fluid where the velocity profile and the fluid flow rate are expressed as \cite{Perkowska_2016}:
\begin{equation}
\label{PL_v}
v(y)=\frac{n}{n+1}\left(-\frac{1}{C}\frac{dp}{dx}\right)^{1/n}\left[\left(\frac{w}{2}\right)^{\frac{n+1}{n}}-y^{\frac{n+1}{n}} \right], \quad Q=\frac{n}{2n+1}2^{-\frac{n+1}{n}}\left(-\frac{1}{C}\frac{dp}{dx}\right)^{1/n}w^\frac{2n+1}{n}.
\end{equation}
The values of $C$ and $n$ are taken from Table \ref{par_table}.
The second benchmark example is the solution to the truncated power-law problem given by the formulae \eqref{v_I_def}--\eqref{v_III_def} and \eqref{Q_1}--\eqref{Q_3}. The channel height, $w$, was set to $10^{-3}\ \text{m}$. The accuracy of computations is assessed by the relative error of the fluid flow rate, $\delta Q$, and the relative error of the fluid velocity, $\delta v$. 
 
\begin{figure}[htb!]
\begin{center}
\includegraphics[scale=0.38]{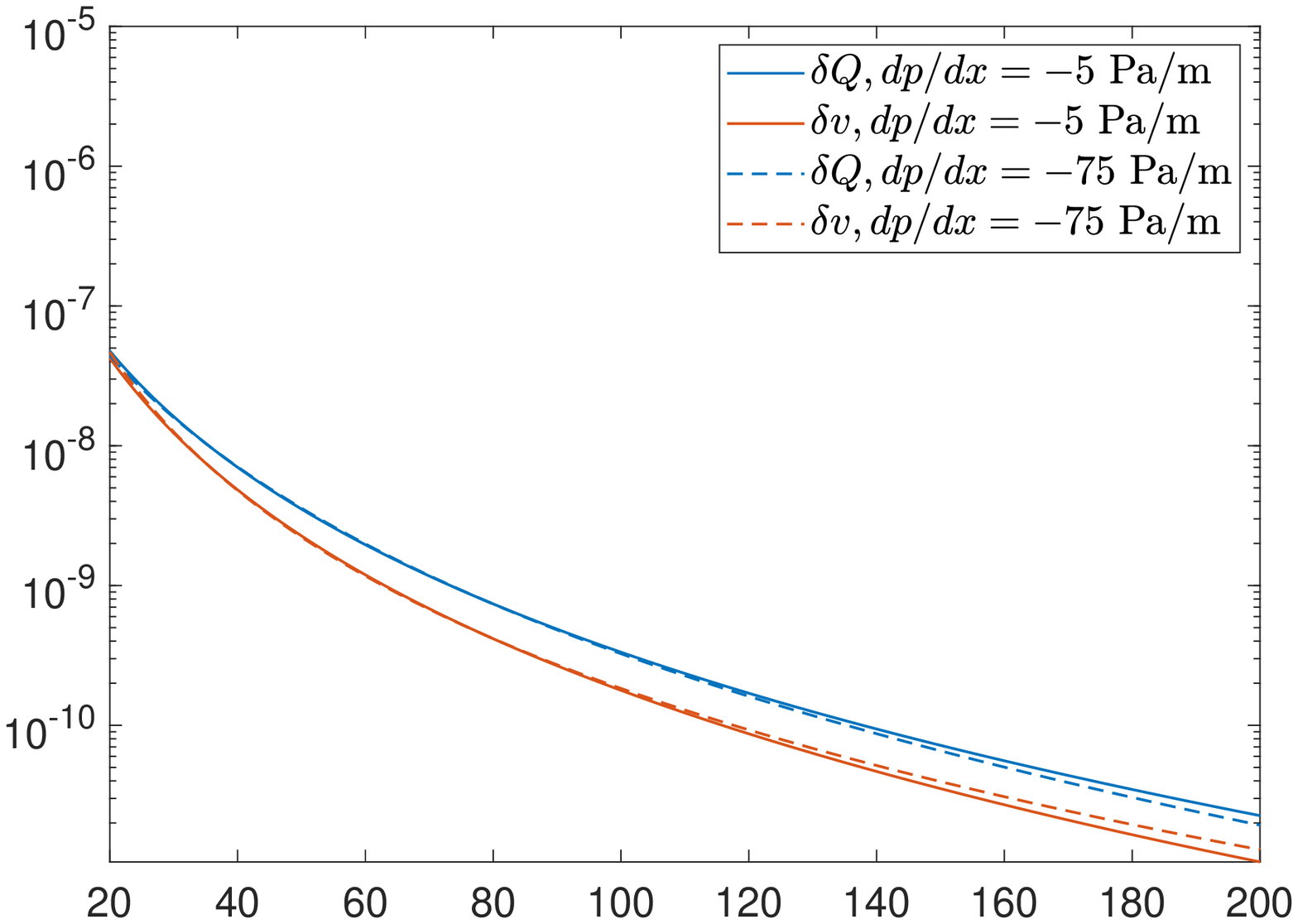}
\hspace{0mm}
\includegraphics[scale=0.38]{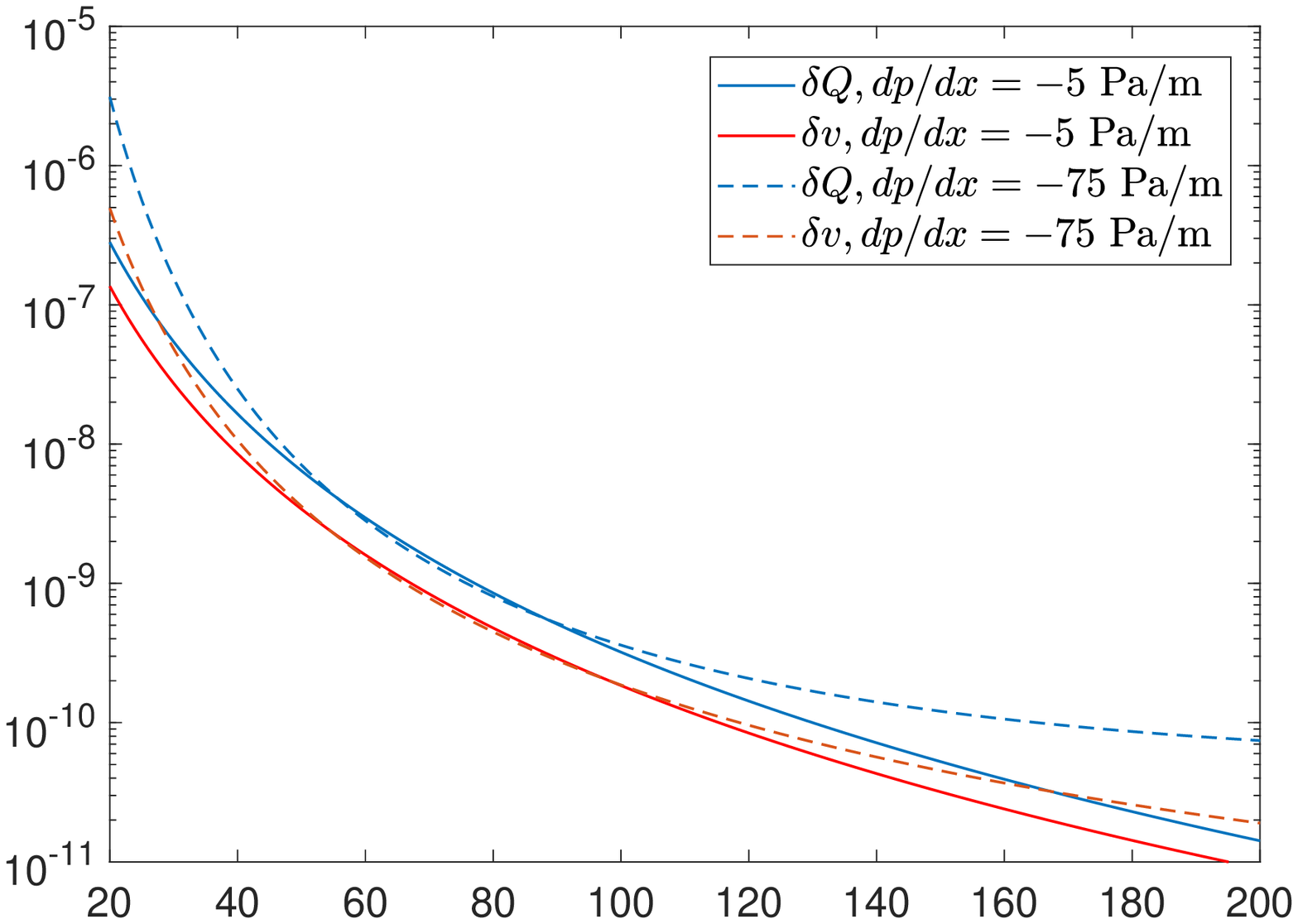}
\put(-338,-5){$M$}
\put(-110,-5){$M$}
\put(-450,140){$\textbf{a)}$}
\put(-220,140){$\textbf{b)}$}
\caption{The relative errors of the fluid flow rate, $\delta Q$, and the maximal relative errors of the fluid velocity, $\delta v$, for: a) the power-law model, b) the truncated power-law model.}
\label{PL_TPL_bledy_N}
\end{center}
\end{figure}

 In Fig. \ref{PL_TPL_bledy_N} the error dependence on the mesh density is shown (for $\delta v$ the maximal value over $y$ is taken) for two values of the pressure gradient: $dp/dx=-5 \ \text{Pa/m}$ and $dp/dx=-75 \ \text{Pa/m}$. As can be seen, the overall accuracy is very good even with only 20 nodal points. The maximal error is of the level of  $10^{-8}$ for the first benchmark and $10^{-6}$ for the second one. With $M=200$ the errors are reduced to the level of $10^{-11}$ regardless of the considered benchmark. 
 
The results show that for a predefined $M$ the accuracy is better for the first benchmark. It stems from the fact that in the second benchmark  the integrand on the right hand side of  \eqref{v_form} is not a smooth function of $y$. Indeed, at the points that correspond to the limiting values of the shear rate ($\dot \gamma_1$ and $\dot \gamma_2$) $\eta_\text{a}$ is only of $C^0$ class. Thus, as cubic spline interpolation of the integrand does not preserve this feature, the overall accuracy is reduced. 

\begin{figure}[htb!]
\begin{center}
\includegraphics[scale=0.38]{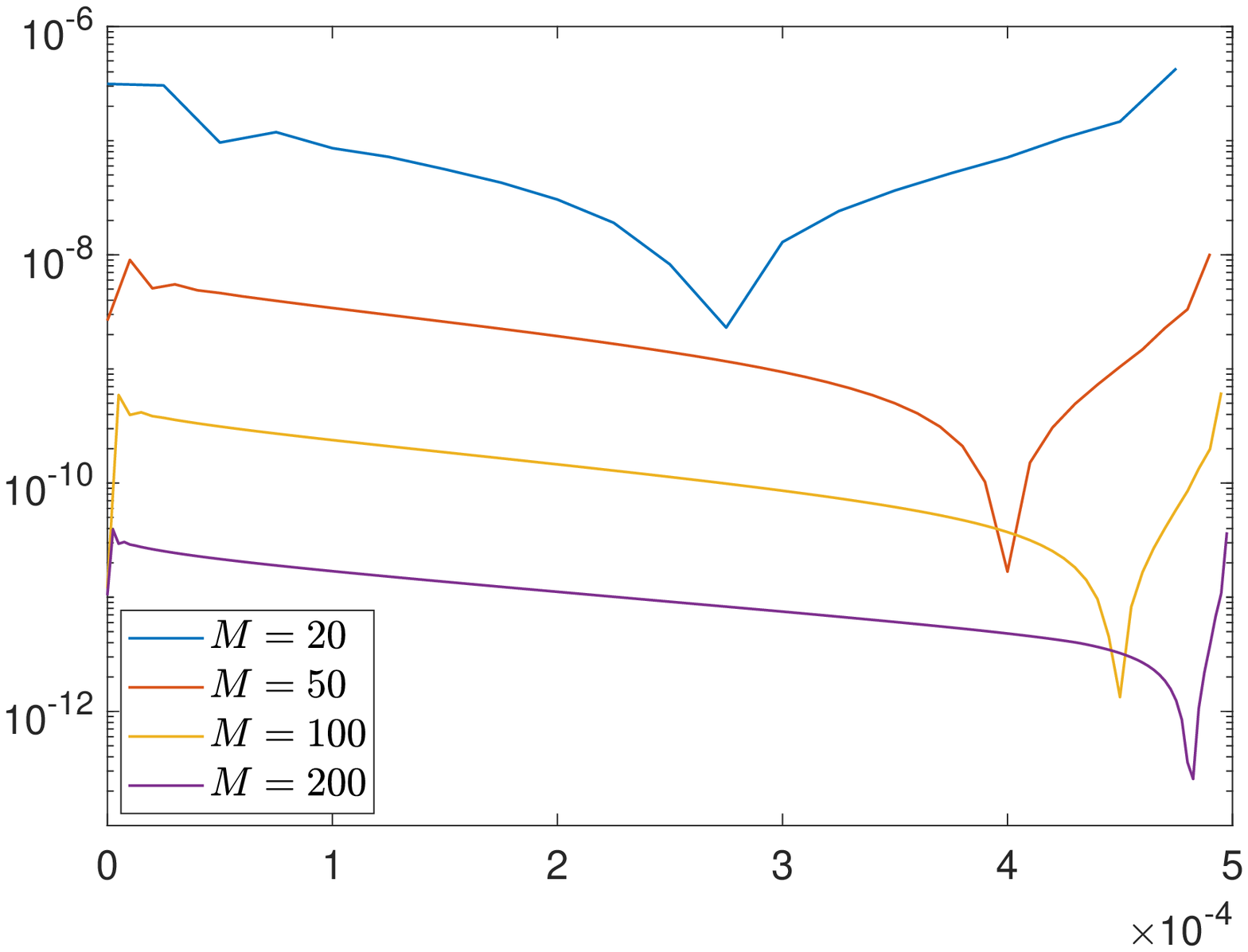}
\hspace{0mm}
\includegraphics[scale=0.38]{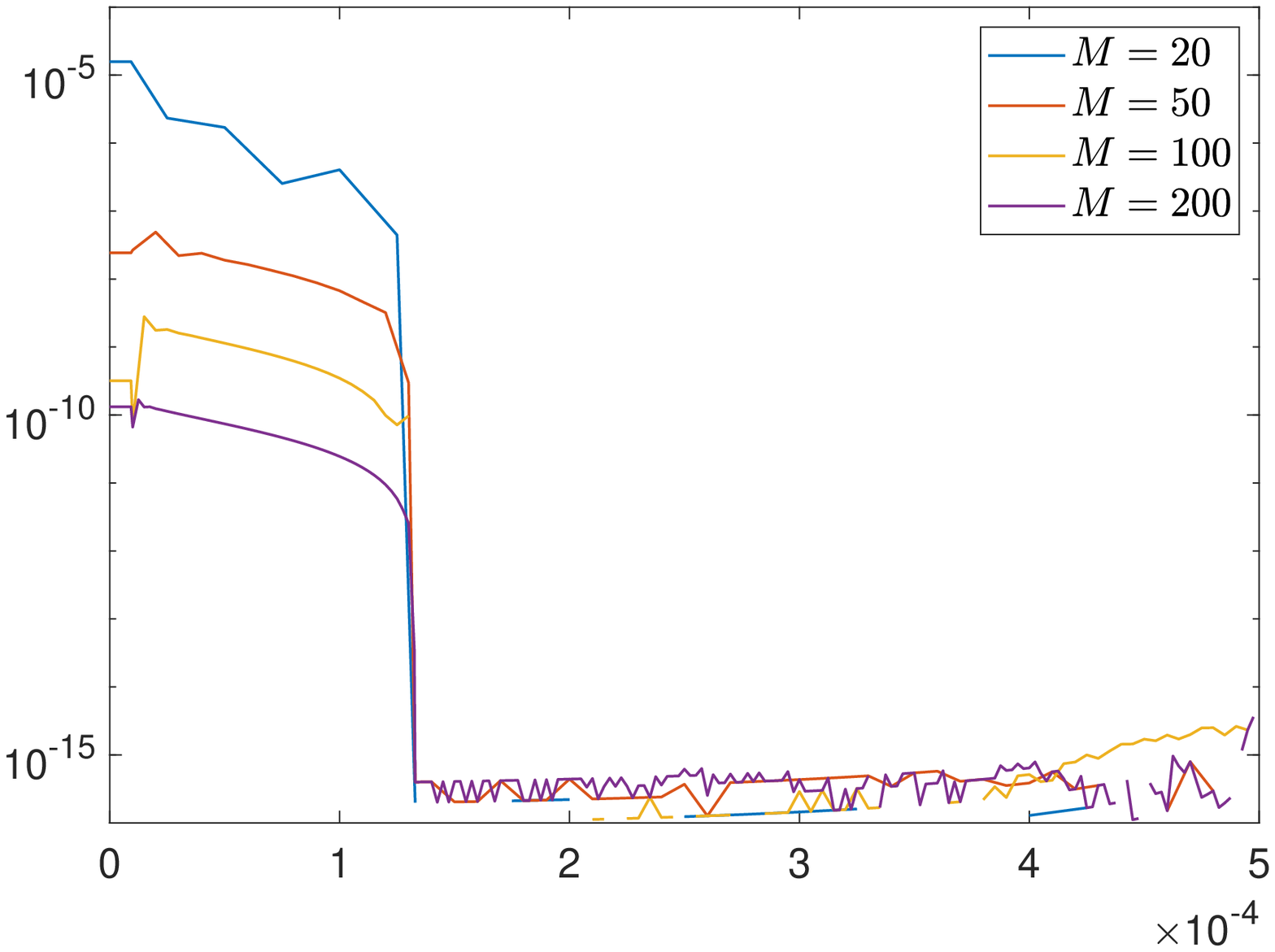}
\put(-338,-5){$y$}
\put(-110,-5){$y$}
\put(-450,80){$\delta v$}
\put(-220,80){$\delta v$}
\put(-450,140){$\textbf{a)}$}
\put(-220,140){$\textbf{b)}$}
\caption{The spatial distributions of the relative error of fluid velocity, $\delta v$, for: a) the power-law model, b) the truncated power-law model. The pressure gradient was $dp/dx=-75 \ \text{Pa/m}$.}
\label{PL_TPL_bledy_y}
\end{center}
\end{figure}

To illustrate this problem we show in Fig. \ref{PL_TPL_bledy_y} the spatial distributions of the velocity errors, $\delta v$. For the second benchmark the limiting values of viscosity were reached for $y\sim 9.26 \cdot 10^{-6}$ ($\dot \gamma_1$) and $y\sim 1.33 \cdot 10^{-4}$ ($\dot \gamma_2$), respectively. Especially in the latter location a sharp error magnification can be observed (Fig. \ref{PL_TPL_bledy_y}b)). Naturally, in the first benchmark no such behavior is present as $\eta_\text{a}$ is  a smooth function of $\dot \gamma$ for that case. Moreover, the same situation holds for the Carreau problem. Thus, we can conclude that the numerical solution obtained by the proposed numerical scheme in the case of Carreau rheology is more accurate than the one computed for the truncated power-law. 

In order to support the last conclusion let us make a comparison of the fluid flow rates, $Q$, obtained for the Carreau law by: i) the proposed numerical scheme, ii) the semi-analytical solution delivered in \cite{Sochi_2015}. Note that the latter assumes solving a respective algebraic equation numerically to find the value of a shear rate at the conduit wall, $\dot \gamma_\text{w}$,  and subsequent substitution to an analytical formula for $Q$. Naturally, the error is generated only at the first stage. When looking for $\dot \gamma_\text{w}$ we set the relative tolerance for this parameter to $10^{-14}$, which defines the accuracy of this reference solution. 

\begin{figure}[htb!]
\begin{center}
\includegraphics[scale=0.4]{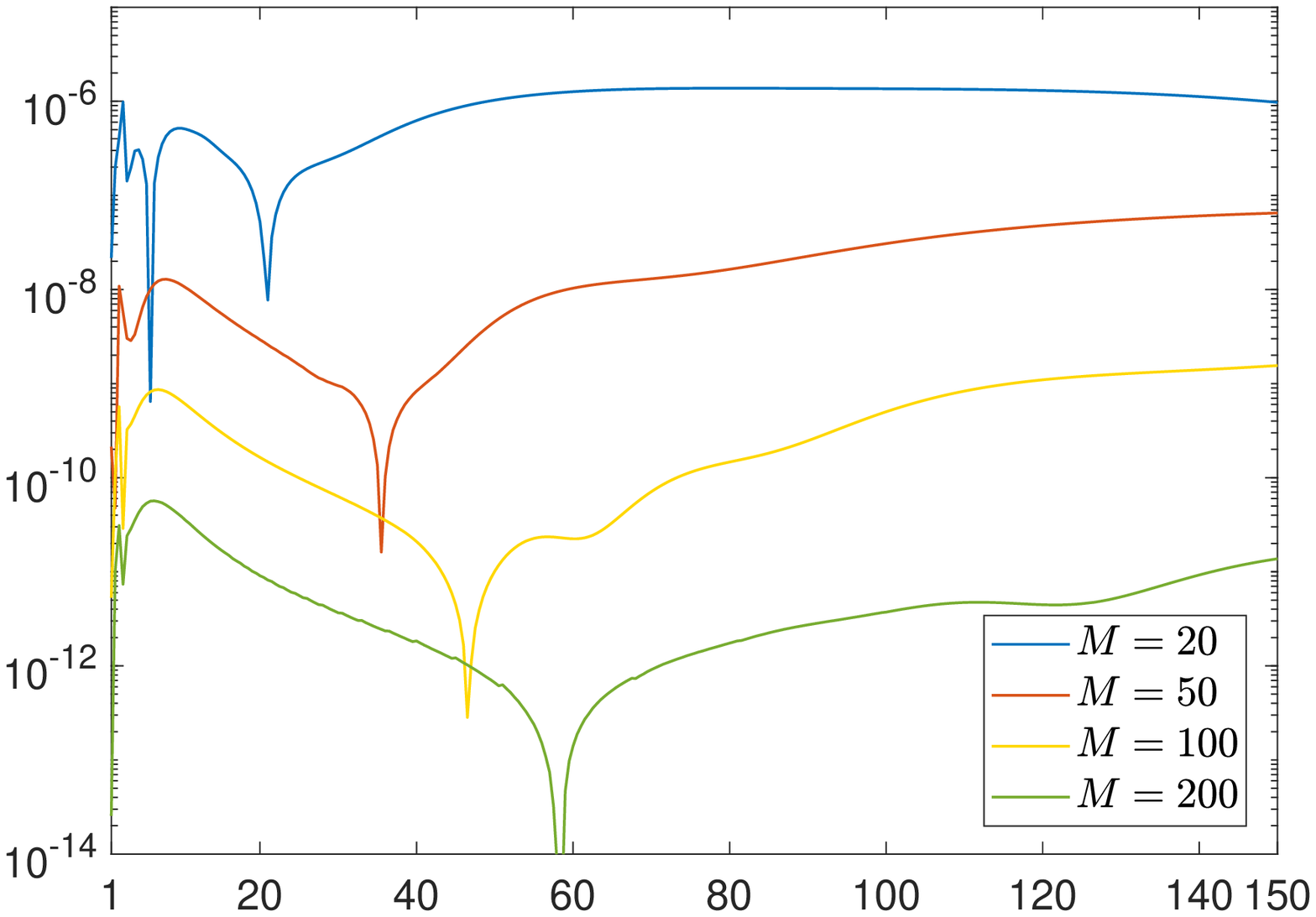}
\put(-230,81){$\delta Q$}
\put(-120,-5){$|dp/dx|$}
\caption{The relative difference between the fluid flow rates, $\delta Q$, computed with the Carreau law  by means of the proposed numerical scheme and the method presented in \cite{Sochi_2015}.}
\label{Sochi_rys}
\end{center}
\end{figure}

The relative differences between the fluid flow rates, $\delta Q$, computed by respective methods are shown in Fig. \ref{Sochi_rys}. Different mesh densities were analyzed for the pressure gradient in the range $1 \ \text{Pa/m}\leq|dp/dx|\leq150  \ \text{Pa/m}$. One can note that the accuracy of the numerical solution delivered by the proposed scheme depends essentially on both, the mesh density and the pressure gradient. The average (over $dp/dx$) values of fluid flow rate deviation, $\delta Q_{av}(M)$, for different $M$ are: $\delta Q_\text{av}(20)=9.9\cdot 10^{-7}$, $\delta Q_\text{av}(50)=2.3\cdot 10^{-8}$, $\delta Q_\text{av}(100)=4.9\cdot 10^{-10}$, $\delta Q_\text{av}(200)=7.0\cdot 10^{-12}$. A comparison of these numbers and the data from Fig. \ref{Sochi_rys} with the characteristics provided in Fig. \ref{PL_TPL_bledy_N}b) allows us to support the claim that the solution  obtained here for the Carreau law is more accurate than the one computed for the truncated power-law benchmark. Thus, the former can be confidently adopted as a reference example when estimating the accuracy of approximate solution introduced in Section \ref{apr_sec}.

\end{document}